\documentclass[onecolumn,draft]{elsart3p}

\bibstyle{elsarticle-num}

\usepackage{amssymb}
\usepackage{footnote}

\journal{arXiv}

\begin{document}
\begin{frontmatter}

\title{Exercises in simplest dynamical random walk, \\
or\, Quantum path integral approach to true diffusion law \\
and 1/f noise  %
of classical particle interacting with ideal gas}

\author{Yuriy E. Kuzovlev}
\ead{kuzovlev@fti.dn.ua}
\address{Donetsk Institute for Physics and Engineering of NASU,
ul.\,R.Luxemburg\,72\,, 83114 Donetsk, Ukraine}


\begin{abstract}
Statistics of classical Hamiltonian random walk of particle %
colliding with atoms of ideal gas is considered from viewpoint %
of earlier suggested exact pseudo-quantum path integral %
representation of the problem, and qualitative agreement %
is demostrated between results of an naturally %
arising simple approximation of %
this integral and results obtained by formally different methods, %
thus in a new fashion showing inevitability of scaleless 1/f-type %
fluctuations in rates of molecular Brownian motions %
and other dynamical transport and evolution processes.
\end{abstract}


\end{frontmatter}

\section{Introduction}

In this paper we turn once again to intriguing questions %
about statistics of walk of a classical particle interacting %
with atoms of thermodynamically equilibrium classical ideal gas. %
This subject already was discussed by us from different points %
of view in \cite{p0806,ig,hs,p1105,p1209} as a principal part %
of more general problem, - considered by %
us in \cite{i1}-\cite{ufn1}, - of molecular ``Brownian'' motion %
in fluids, molecular and electronic transport processes in  %
various many-particle dynamical systems,  %
and related fundamental 1/f noises %
\footnote{\,
Short partial reviews of corresponding ideas and results can be %
found in \cite{i1,i2,pr157,ufn,p0802,tmf,p1008, %
p1203,ufn1} (see also introductory and discussion sections in %
\cite{i3,kmg,p1007,e2,mang,p1207,p1302}).}\,. %
In \cite{p0806} we suggested an original exact representation %
of characteristic function and thus probability distribution of %
displacement (path) of the ``Brownian particle'' in terms %
of effective quantum transition amplitudes or equivalent %
holomorphic path integrals (on them see e.g. \cite{sf,cd}). %
Now, we  shall discern these mathematical objects more %
carefully and try some their analytical approximations %
which visually reveal their main physical contents. %
Namely, the fact that the ``Brownian particle'' (BP) %
has no definite diffusion and drift rates %
(diffusivity and mobility) but walks %
as if these quantities undergo 1/f\,-type %
low-frequency fluctuations.

Generally, all transport processes in many-particle dynamical %
(Hamiltonian)  systems are statistical  %
analogues of Brownian motion and all also have no %
certain (a priori predictable time-averaged) %
rate. In essence, this was foreseen by N.\,Krylov %
as long ago as at end of the forties %
\cite{kr}. In the eighties, this truth %
was realized independently \cite{i1,pr157,bk1,bk2,ufn,pr195} %
along with its direct relation to the widely %
observed 1/f-noise (``flicker noise''). %
Unfortunately, today's priests of %
statistical mechanics still stay %
constrained by prejudices which took beginning %
in the works by L.\,Boltzmann on gas kinetics and were %
hallowed by his authority %
although later disclosed in \cite{kr}. %
Such the situation notably damages theoreticians' %
``human rights'' in statistical physics %
and enforces applied science and engineering to search %
for 1/f-noise sources in particular physical objects, - %
like e.g. ``charge traps'' or some %
(usually mysterious) ``two-level fluctuators'', - %
instead of general statistical properties of %
transport processes produced by deterministic %
Hamiltonian dynamics. %

This explains our interest in gases and even %
so ``simple'' system as a particle in ideal gas: %
clearly, if these systems create 1/f-noise %
then its existence does not require ``traps'', %
``slow fluctuators'' and any ``long memory times'' %
at all. Of course,  charge traps and two-level %
defect structures in solids may take part in %
formation of 1/f-noise but not necessarily %
as its sources. %
The true universal origin of 1/f-noise is that %
complex enough (``mixing'') many-particle dynamics %
constantly forgets its past and therefore  %
does not control relative frequencies of %
elementary kinetic events (various ``collisions'', %
``scatterings'', ``tunnel transitions'', %
``trappings'' and ``de-trappings'', ``reactions'', %
etc.). %
Hence, to assume for them \,{\it a priory} settled ``probabilities %
per unit time'' means to lose a good few of actual %
chances, namely, lose 1/f-noise. %
That is just what Boltzmann-like kinetic %
models always do %
\footnote{\, %
At the expense of logical and mathematical %
self-consistency of theory, as it was especially highlighted in %
\cite{p1203} and in last section of %
\cite{p0806}.}\,. %

By these reasons, unprejudiced investigations of actual gas %
kinetics, - based on exact evolution equations for %
one-, two-, three- and all other many-particle %
probability distribution functions, - is %
of key importance for kinetics as the whole %
including phenomenon of 1/f-noise.
Since the latter by its physical nature  %
manifests memoryless features of %
underlying dynamics, its formal  %
extraction from exact equations of dynamics is %
very fine mathematical operation. %
Therefore any its correct enough intuitively %
transparent simplification %
would be very useful. Below, we shall  %
make some steps in this direction, %
at that using definitions, designations and formulae %
from \cite{p0806} (see also \cite{p0804} or \cite{tmf}).

\section{Pseudo-quantum representations of diffusion law}

{\bf 1}.\,\, %
In comparison with \cite{p0806}, here, - like %
in \cite{p1105,p1209}, - we shall consider more general %
case of possibly non-zero external force \,$f$\, %
applied to the ``Brownian''' particle (BP) after initial time moment %
\,$t=0$\,. At \,$t=0$\,  the BP is settled %
to start from point \,$R(0)=0$\, possessing Maxwellian %
velocity distribution and being surrounded by %
equilibrium (ideal) gas. %

Now, all the correlation functions (CF) \,$V_n$\, acquire %
additional argument \,$f$\,, and evolution operator %
in the exact evolution equation %
\begin{equation}
\begin{array}{l}
\frac {\partial \mathcal{V}(t)}{\partial t}\,=\,\widehat{\pounds}\,
\mathcal{V}(t)\,\, \label{fev1}
\end{array}
\end{equation}
for their generating functional %
\,$\mathcal{V}(t) =\mathcal{V}\{t,ik,P,\psi\,;\nu,\,f\}$\, looks as
\begin{equation}
\begin{array}{l}
\widehat{\pounds}\,=\,\,i{\bf k}\cdot{\bf V}\,-\, %
f\cdot\frac {\partial}{\partial P}\,+ %
\int \psi(x)\left[\,({\bf V}-{\bf v})\cdot\frac {\partial}{\partial
\rho}\,+\,\Phi^{\prime}(\rho)\cdot\left(\frac {\partial}{\partial
{\bf p}}-\frac {\partial}{\partial {\bf P}}\right)\,\right]\frac
{\delta }{\delta
\psi(x)}\,-\,\label{mop}\\
-\,\frac {\partial }{\partial {\bf P}} \int
\Phi^{\prime}(\rho)\,\frac {\delta }{\delta
\psi(x)}\,+\,\nu\left[\int G_m({\bf
p})\,E^{\,\prime}(\rho)\,\psi(x)\,\right]\cdot\left({\bf V}+T\frac
{\partial }{\partial {\bf P}}\right)\,\,\, 
\end{array}
\end{equation}
or, equivalently,
\begin{equation}
\begin{array}{l}
\widehat{\pounds}\,=\,i{\bf k}\cdot{\bf V}\,-\, %
f\cdot\frac {\partial}{\partial P}\,+\, \label{mop_}\\
+\int [\,1+\psi(x)\,]\, %
\left[\,({\bf V}-{\bf v})\cdot\frac {\partial}{\partial %
\rho}\,+\,\Phi^{\prime}(\rho)\cdot %
\left(\frac {\partial}{\partial %
{\bf p}}-\frac {\partial}{\partial {\bf P}}\right)\,\right]\, %
\left[\,\frac {\delta }{\delta \psi(x)}\,+  %
\nu\, G_m({\bf p})\,E(\rho)\,\right]\,\,
\end{array}
\end{equation}
The same as before (equilibrium) initial condition %
to Eq.\ref{fev1} is %
\,$\mathcal{V}\{t=0\}\,=\,G_M({\bf P})$\, %
(Eq.30 from \cite{p0806}). %
Recall that \,$k$\, is wave vector conjugated with %
BP's coordinate \,$R$\,,\, \,$\nu$\, is mean gas density,\, %
\,$x=\{\rho,p\}$\, enumerate phase space points of gas atoms, - %
with \,$\rho=r-R$\, denoting distance of an atom from BP, - %
\,$\Phi(\rho)$\, is BP-atom interaction potential,\,  %
\,$E(\rho)=\exp{[-\Phi(\rho)/T]}$\,,\, and %
\,$G_m(p)$\, and \,$G_M(P)$\, are Maxwelliam distributions %
for momenta of gas atoms and BP, respectively, %
with \,$p=mv$\, and \,$P=MV$\, (please, see \cite{p0806} %
for details).

Introducing, as in \cite{p0806}, auxiliary %
boson creation and annihilation operators: %
\begin{equation}
\begin{array}{l}
{\bf A}^\dag\,\equiv\,-\,\sqrt{TM}\,\frac {\partial }{\partial {\bf
P}}\,\,\,\,,\,\,\,\,\,\,{\bf A}\,\equiv\,\sqrt{\frac MT}\,\left({\bf
V}+T\frac {\partial }{\partial {\bf
P}}\right)\,\,\,,\,\,\,\,\,\,A_{\alpha}A^\dag_{\beta}
-A^\dag_{\beta}A_{\alpha}\,=\,%
\delta_{\alpha\beta}\,\,\,, \label{aa}
\end{array}
\end{equation}
\begin{equation}
\begin{array}{l}
a^\dag(x)\,\equiv\,c(x)\,\psi(x)\,\,\,\,,\,\,\,\,\,\,a(x)\,\equiv\,
c^{-1}(x)\,\frac {\delta }{\delta \psi(x)}\,\,\,,
\,\,\,\,\,\,\,\,\,a(x)a^\dag(y)-a^\dag(y)a(x)\,=
\,\delta(x-y)\,\,\,,\label{ax}
\end{array}
\end{equation}
with \,$c(x)\,=\,\sqrt{\nu\,G_m({\bf p})\,E(\rho)}$\,, %
and operators
\begin{equation}
\begin{array}{l}
\widehat{\pounds}_1\,=\, %
\widehat{\pounds}_1(ik,f)\,=\, %
\,u_0\,i{\bf k}\cdot{\bf A} + %
u_0\,\left( i{\bf k} + \frac {{\bf f}}T\right)\cdot %
{\bf A}^\dag\,\,\,,\,\label{mop1}
\end{array}
\end{equation}
\begin{equation}
\begin{array}{l}
\widehat{\pounds}_2\,=\,\,\int a^\dag(x)\left[-{\bf v}\cdot\frac
{\partial}{\partial \rho}\,+\,\Phi^{\prime}(\rho)\cdot\frac
{\partial}{\partial {\bf p}}\,\right]a(x)\,+\,u_0\int c(x)\,\frac
{\Phi^{\prime}(\rho)}{T}\cdot[\,a(x){\bf A}^\dag\,-\,a^\dag(x) {\bf
A}\,]\,\,\,,\label{mop2}
\end{array}
\end{equation}
\begin{equation}
\begin{array}{l}
\widehat{\pounds}_3\,=\,u_0\int a^\dag(x)\left[\,({\bf A}^\dag+{\bf
A})\cdot\frac {\partial}{\partial \rho}\,+\,({\bf A}^\dag-{\bf
A})\cdot\frac {\Phi^{\prime}(\rho)}{2T}\, %
\right]a(x)\,\,, \, \label{mop3}
\end{array}
\end{equation}
\[
\widehat{\pounds}(ik,f)\,=\,\widehat{\pounds}_1(ik,f)\,+
\,\widehat{\pounds}_2\,+\,\widehat{\pounds}_3\,\,\,, 
\]
with $\,u_0=\sqrt{T/M}\,$, %
we represent the characteristic function of BP's %
displacement (path) in the form of ``vacuum average'': %
\begin{equation}
V_0(t,i{\bf k};\,\nu,f)\,=\, %
\langle\,e^{ik\cdot[R(t)-R(0)]}\,\rangle \,=\, %
\langle
0|\,\,e^{\,t\widehat{\pounds}(ik,f)}\,\,|0\rangle \,\,\label{fa1}
\end{equation}
At that, the vacuum state corresponds to complete  %
statistical equilibrium in \,$\{\rho,p,P\}$\,-space %
(absence of any BP-gas statistical correlations %
except purely equilibrium ones). 

\,\,\,

{\bf 2}.\,\, %
We may allow the wave vector \,$k$\,,  %
as well as the external force, %
to be arbitrary probe function of time, \,$k=k(t)$\,. %
Then instead of (\ref{fa1}) we have to write
\begin{equation}
V_0(t)\,=\, %
\langle\,\exp{[ \int_0^t ik(t^\prime)\, %
V(t^\prime)\,dt^\prime\,]} \,\rangle \,=\, %
\langle 0|\,\,\overleftarrow{\exp}{\,[ %
\int_0^t \widehat{\pounds}(ik(t^\prime), %
f(t^\prime))\, dt^\prime\,]}\,|0\rangle \,\,\label{fa1_}
\end{equation}

\section{Path integral %
representations of diffusion law %
and its stochastic formulation}

{\bf 1}.\,\,
Reducing standardly \cite{sf,cd} the ``vacuum-vacuum %
quantum transitions amplitudes'' (\ref{fa1}) and (\ref{fa1_}) %
to the holomorphic, or coherent-state,  path integrals, %
one has
\begin{equation}
\begin{array}{l}
V_0(t)\,= \, \int %
\exp{\left\{\int^t_0 \left[ \frac 12\,
(\,\dot{\mathcal{A}}^*\mathcal{A}\,-\,\mathcal{A}^*\dot{\mathcal{A}}\,)
\,+\,\widehat{\pounds}(ik,f,\mathcal{A}^*,\mathcal{A})
\right]d\xi\right\}}\,\prod_{\xi} \frac
{d\mathcal{A}^*d\mathcal{A}}{2\pi i}\,\,, \label{fa2}
\end{array}
\end{equation}
with
$\,\mathcal{A}^*=\mathcal{A}^*(\xi)= \{a^*(x,\xi),{\bf A}^*(\xi)\}  %
\,$ and $\,\mathcal{A}=\mathcal{A}(\xi) =\{a(x,\xi),{\bf A}(\xi)\}\,$ %
being complex variables, in place of boson operators, %
and edge conditions\, $\,\mathcal{A}^*(\xi=t)=0\,$,\,
$\,\mathcal{A}(\xi=0)=0\,$\,.

\,\,\,

{\bf 2}.\,\, %
The exponential in Eq.\ref{fa2} is quadratic in respect to %
the gas variables $\,\{a^*(x,\xi),\,a(x,\xi)\}\,$\, %
as well as BP variables $\,\{A^*(x,\xi),\,A(x,\xi)\}\,$\, %
separately, therefore integration over one of them can be %
performed exactly. Integration over gas yields

\begin{equation}
\begin{array}{l}
V_0(t,i{\bf k};\,\nu,f)\,=\,\int \, \exp{\left\{\int^t_0 \left[ \frac 12\,
(\dot{{\bf A}}^*\cdot{\bf A}-{\bf A}^*\cdot\dot{{\bf A}}) \,+\, %
i{\bf k}\cdot{\bf A}\,u_0 + %
(i{\bf k} + {\bf f}/T)\cdot {\bf A}^*\,u_0\,
\right]d\xi\right\}}\,\times\,\label{fa3}\\
\,\,\,\,\,\,\,\,\,\,\,\,\,\,\,\,\,\,\,\,\,\,\,\,\,\,\,\,\,\,\,\,
\times\,\,\exp{\left[-\int_{t>\,\xi_1>\,\xi_2>\,0} A_{\alpha}^*(\xi_1)
\,\,\mathbb{G}_{\alpha\beta}\{\xi_1,\xi_2,\,{\bf A}^*,{\bf A}\}\,\,
A_{\beta}(\xi_2)\,\,d\xi_2\,d\xi_1\right]}\, %
]\,\prod_{\xi} \frac {d{\bf A}^*d{\bf
A}}{(2\pi i)^d}\,\,=\,\,
\end{array}
\end{equation}
\begin{equation}
\begin{array}{l}
=\,\int \, \exp{\left\{\int^t_0 \left[ \frac
12\, (\dot{{\bf A}}^*\cdot{\bf A}-{\bf A}^*\cdot\dot{{\bf A}})
\,+\, i{\bf k}\cdot{\bf A}\,u_0 + %
(i{\bf k} + {\bf f}/T)\cdot {\bf A}^*\,u_0\, %
\right]d\xi\right\}}\,\times\,\label{fa4}\\\,\,\,\,\,\,\,\,\,\,\,\,\,\,\,
\,\,\,\,\,\,\,\,\,\,\,\,\,\,\,\,\,\,\times\,\,\exp{\left\{\int c(x)
\left(\overleftarrow\exp{\left[ \int^t_0 \widehat\Lambda({\bf
A}^*,{\bf A})\,d\xi\right]}-1\right)c(x)\right\}} \,\,\prod_{\xi}
\frac {d{\bf A}^*d{\bf A}}{(2\pi i)^d}\,\,\,,
\end{array}
\end{equation}
with edge conditions\, $\,{\bf A}^*(t)=0\,$,\, $\,{\bf A}(0)=0\,$\,,
summation over repeated indices, and %
 $\,\int\,...\,=\int\int\,...\,d{\bf
p}\,d\rho\,=\int \dots dx\,$. %
The kernel in Eq.\ref{fa3} is
\begin{equation}
\begin{array}{l}
\mathbb{G}_{\alpha\beta}\{\xi_1,\xi_2,\,{\bf A}^*,{\bf A}\}\,=\,
\frac {u_0^2}{T^2}\int c(x)\,\Phi^{\prime}_{\alpha}(\rho)\,\,
\overleftarrow\exp{\left[ \int^{\xi_1}_{\xi_2} \widehat\Lambda({\bf
A}^*(\xi),{\bf
A}(\xi))\,d\xi\right]}\,\,\Phi_{\beta}^{\prime}(\rho)\,c(x)
\,\,\,,\label{kern}
\end{array}
\end{equation}
and\, $\,\widehat\Lambda({\bf A}^*,{\bf A})\,$\, %
in Eqs.\ref{fa3}-\ref{fa4} means Liouville-like operator
\begin{equation}
\begin{array}{l}
\widehat\Lambda({\bf A}^*,{\bf A})\,=\, -\,{\bf v}\cdot\frac
{\partial}{\partial \rho}\,+\,\Phi^{\prime}(\rho)\cdot\frac
{\partial}{\partial {\bf p}}\,+\,\,u_0\left[({\bf A}^* +{\bf
A})\cdot\frac {\partial}{\partial \rho}\,+\,({\bf A}^*-{\bf
A})\cdot\frac {\Phi^{\prime}(\rho)}{2T}\right]\,\label{L}
\end{array}
\end{equation}


\,\,\,

{\bf 3}.\,\,%
Integration in (\ref{fa2}) firstly over %
BP's variables $\,{\bf A}\,$ and $\,{\bf A}^*\,$ %
leads to holomorphic path integral
\begin{equation}
\begin{array}{l}
V_0(t,i{\bf k};\,\nu,f)\,=\, %
\int \exp{\left\{\int^t_0 \int\left[ \frac
12\, (\,\dot{a}^*a\,-\,a^*\dot{a}\,)\,+\,a^*\,\widehat\Lambda_0\, a
\,\right]dx\,d\xi\right\}}\,\times\,\label{fa5}\\
\times\,\,\exp{\left\{u_0^2\int_0^t d\xi_1\int_0^{\xi_1}d\xi_2
\,\,[\,i{\bf k}\,+\,{\bf K}^*(a^*(\xi_1),a(\xi_1))\,]\cdot %
\left[\,i{\bf k} + \frac {{\bf f}}T\,+\, %
{\bf K}(a^*(\xi_2),a(\xi_2))\,\right]\, %
\right\}} \, \prod_{x,\,\xi}
\frac {da^*da}{2\pi i}\,\,\,,\,\label{fa5}
\end{array}
\end{equation}
with edge conditions\, $\,a^*(x,\xi=t)=0\,$, $\,a(x,\xi=0)=0\,$,\, %
functionals
\[
\begin{array}{l}
{\bf K}^*(a^*,a)\,=\,-\int c(x)\,\frac
{\Phi^{\prime}(\rho)}{T}\,\,a^*(x)\,dx\,+\,\int a^*(x)\left[\,\frac
{\partial}{\partial \rho}-\frac
{\Phi^{\prime}(\rho)}{2T}\right]\,a(x)\,dx\,\,\,,
\end{array}
\]
\[
\begin{array}{l}
{\bf K}(a^*,a)\,=\,\int c(x)\,\frac
{\Phi^{\prime}(\rho)}{T}\,\,a(x)\,dx\,+\,\int a^*(x)\left[\,\frac
{\partial}{\partial \rho}+\frac
{\Phi^{\prime}(\rho)}{2T}\right]\,a(x)\,dx\,\,\,,
\end{array}
\]
and Liouville operator
\[
\begin{array}{l}
\widehat\Lambda_0\,=\, -\,{\bf v}\cdot\frac {\partial}{\partial
\rho}\,+\,\Phi^{\prime}(\rho)\cdot\frac {\partial}{\partial {\bf
p}}\,\,\label{L0}
\end{array}
\]
of an atom interacting with BP as if the latter  %
was immovable.

\,\,\,

{\bf 4}.\,\, %
Let us introduce, as in Sec.6 in \cite{p0806}, %
the ``seed'' Gaussian measure %
\,$\mathcal{M}_0\{\mathcal{A}^*,\mathcal{A}\}$\, %
in space of holomorphic paths by %
\[
\begin{array}{l}
d\mathcal{M}_0\{\mathcal{A}^*,\mathcal{A}\}\,=\, %
\exp{\left\{\int \frac 12\,
(\,\dot{\mathcal{A}}^*\mathcal{A}\,-\,\mathcal{A}^*\dot{\mathcal{A}}\,)
\,d\xi\,\right\}} %
\,\prod_{\xi} \frac {d\mathcal{A}^*d\mathcal{A}}{2\pi
i}\,\,
\end{array}
\]
In comparison with \cite{p0806}, we added subscript %
``0'' to underline the ``seed'' character of this measure. %
It is completely determined by its %
characteristic functional
\begin{equation}
\begin{array}{l}
\left\langle\, \exp{\left\{\,\int_0^t
[\,b(\xi)\cdot\mathcal{A}^*(\xi)+b^*(\xi)\cdot\mathcal{A}(\xi)\,]\,d\xi
\,\right\}}\,\right\rangle_0\,\equiv\, \\ %
\equiv\,
\int \exp{\left\{\,\int_0^t
[\,b(\xi)\cdot\mathcal{A}^*(\xi)+b^*(\xi)\cdot\mathcal{A}(\xi)\,]\,d\xi
\,\right\}}\, d\mathcal{M}_0\{\mathcal{A}^*,\mathcal{A}\} %
\,=\, \\ \,=\, %
\exp{\left[\,\int_0^t dt^{\,\prime}
\int_0^{t^{\,\prime}}dt^{\,\prime\prime}\,\,b^*(t^{\,\prime})\cdot
b(t^{\,\prime\prime})\,\right]}\,\,\,\label{cfa}
\end{array}
\end{equation}
or, equivalently, by corresponding %
pair correlators
\begin{equation}
\begin{array}{l}
\langle\,\mathcal{A}_{\alpha}(t_1)\,\mathcal{A}_{\beta}(t_2)\, %
\rangle_0\,=
\,\langle\,\mathcal{A}^*_{\alpha}(t_1)\,\mathcal{A}^*_{\beta}(t_2)\,
\rangle_0\,=\,0\,\,\,,
\,\,\,\,\,\,\langle\,\mathcal{A}_{\alpha}(t_1)\,
\mathcal{A}^*_{\beta}(t_2)\,\rangle_0\,
=\,\delta_{\alpha\beta}\,\Theta(t_1-t_2)\,\,\,,\label{sm}
\end{array}
\end{equation}
where $\,\Theta(t)\,$ denotes the Heaviside function. %

Thus, for  arbitrary functional
$\,\mathbb{F}\{ \mathcal{A}^*(\xi),\mathcal{A}(\xi)\}\,$ %
we may write
\[
\int \exp{\left\{\int \frac 12\,
(\,\dot{\mathcal{A}}^*\mathcal{A}\,-\,\mathcal{A}^*\dot{\mathcal{A}}\,)
\,d\xi\,+\,\mathbb{F}\{\mathcal{A}^*(\xi),\mathcal{A}(\xi)\}\right\}}
\,\prod_{\xi} \frac {d\mathcal{A}^*d\mathcal{A}}{2\pi
i}\,=\, %
\langle\, \exp{\,\mathbb{F}\{\mathcal{A}^*(\xi), %
\mathcal{A}(\xi)\}}\, \rangle_0\,\, 
\]
In particular, consequently,
\begin{equation}
\begin{array}{l}
V_0(t,i{\bf k};\,\nu,f)\,=\,\langle
0|\,\,e^{\,t\widehat\pounds(\mathcal{A}^\dag,\,\mathcal{A})}\,
\,|0\rangle\,=\,\left\langle \exp{\int^t_0
\widehat\pounds(\mathcal{A}^*(\xi),\mathcal{A}(\xi))\, %
d\xi}\right\rangle_0 \,\,\label{fa6}
\end{array}
\end{equation}
with function\, %
$\,\widehat\pounds(\mathcal{A}^*,\,\mathcal{A})\,$ %
(``normal symbol'' \cite{sf,cd} of operator %
$\,\widehat\pounds(\mathcal{A}^\dag,\,\mathcal{A})\,$) %
defined by (\ref{mop1})-(\ref{mop3}), and
\begin{equation}
\begin{array}{l}
V_0(t,i{\bf k};\nu,f)\,=\,\langle\, %
\exp\,\{\int^t_0 %
u_0\,[\,i{\bf k}\cdot {\bf A}(\xi)+ %
(i{\bf k} + {\bf f}/T ) %
\cdot {\bf A}^*(\xi)\, ]\,d\xi\, %
-\, \\ \,\, - %
\int_{t>\,\xi_1>}\int_{\xi_2>\,0} {\bf A}^*(\xi_1)
\cdot\mathbb{G}\{\xi_1,\xi_2,{\bf A}^*,{\bf A}\} %
\cdot {\bf A}(\xi_2) \,d\xi_2\,d\xi_1 %
\,\}\,\rangle_0\, \label{sv}
\end{array}
\end{equation}
with Gaussian random processes $\,{\bf
A}^*(t)\,$ and $\,{\bf A}(t)\,$ obeying (\ref{sm}). %

\,\,\,

{\bf 5}.\,\,
Notice that formula (\ref{fa4}), - and hence %
(\ref{fa3}) and (\ref{sv}), - %
is in fact exactly the same as, firstly, visually different %
path integral expressed by Eqs.42-44 from \cite{sr}, %
and, secondly, the ``dynamical virial expansion''  %
of BP's diffusion law presented in \cite{p1209} by Eq.7 or %
Eq.11 (at \,$s=0$) and derived directly from non-equilibrium %
partition function.

\section{The interaction kernel and %
fixed relaxation rate approximation} %
{\bf 1}.\,\,
Obviously, properties of BP's characteristic function %
(\ref{sv}) and thus its diffusion law %
\[
\begin{array}{l}
V_0(t,\Delta R;\nu,f)\,=\, %
\int e^{-ik\cdot \Delta R}\, %
V_0(t,ik;\nu,f)\, %
\frac {d^3k}{(2\pi)^3}\,
\end{array}
\]
are fully determined by properties  %
of the kernel  \,$\mathbb{G}$\,. %
From \,$\mathbb{G}$\,'s definition (\ref{kern})-(\ref{L}) one can see %
that it describes relaxation and simultaneously thermal agitation %
of BP's momentum and velocity due to BP's %
interactions with atoms. Below, for certainty %
and simplicity, we shall think of these %
interactions as ``collisions'', thus assuming the potential %
\,$\Phi(\rho)$\, repulsive and short-range enough %
(so that atoms can not form ``bound states'' with BP). %
Let us discuss the kernel in corresponding terms.

Dependence of the kernel %
\,$\mathbb{G}\{\xi_1,\xi_2,{\bf A}^*,{\bf A}\}$\, %
on BP's boson variables %
reflects three physical factors.
First of them is that results of any BP-atom collision %
are sensitive to non-inertial character of %
BP's motion during this collision and %
thus to atom-BP mass ratio \,$m/M$\,.
In Eq.\ref{sv} this is described by %
\,${\bf A}(\xi^\prime)$ - \,${\bf A}^*(\xi^{\prime\prime})$\,  %
correlations ``inside the kernel'' %
(at \,$\xi_1>\xi^\prime >\xi^{\prime\prime}>\xi_2$\,).

Second factor is that effect of any particular collision %
depend on degree of current non-equilibrium of the system and thus on %
pre-history of its evolution.
This is described by cross-correlations between %
the kernel's inner variables and outer variables what %
``frame'' it (\,${\bf A}(\xi_1)$\, %
and \,${\bf A}^*(\xi_2)$\, in Eq.\ref{sv}).

Third factor is that contribution %
of any particular collision %
to summary BP's path  \,$\Delta R(t)=R(t)-R(0)$\, %
interferes with all other, previous and next, collisions %
(moreover, previous ones determine very  occurance %
of present one).
In Eq.\ref{sv} this is described %
by inter-correlations of \,${\bf A}(\xi^\prime)$\, %
and \,${\bf A}^*(\xi^{\prime\prime})$\, belonging %
to different copies of the kernel (for instance, %
between \,${\bf A}(\xi^\prime)$\, from %
\,$\mathbb{G}\{\xi_1,\xi_2,{\bf A}^*,{\bf A}\}$\, and %
\,${\bf A}^*(\xi^{\prime\prime})$\, from %
\,$\mathbb{G}\{\xi_3,\xi_4,{\bf A}^*,{\bf A}\}$\, %
at \,$\xi_2>\xi_3$\,).

It is important to realize that usual %
(``Boltzmannian'') kinetics deals with the first %
and partly with the second %
factors only and loses the third  %
since avoids consideration of actual variety of %
spatial-temporal clusters of collisions.

\,\,\,

{\bf 2}.\,\,
Clearly, to take into account the first of these factors, %
it is quite sufficient to approximate the exact functional %
\,$A^*,\,A$\,-dependent kernel \,$\mathbb{G}$\, %
by its average value, that is to make replacement
\begin{equation}
\begin{array}{l}
\mathbb{G}_{\alpha\beta}\{\xi_1,\xi_2,\,{\bf A}^*,{\bf A}\}\, %
\Rightarrow\, %
\mathcal{G}_{\alpha\beta}(\xi_1-\xi_2)\,\equiv\, %
\langle\, %
\mathbb{G}_{\alpha\beta}\{\xi_1,\xi_2,\,{\bf A}^*,{\bf A}\}\, %
\rangle_0\, =\, \label{rep}\\ =\,
\frac {u_0^2}{T^2}\int c(x)\,\Phi^{\prime}_{\alpha}(\rho)\, %
\left\langle\, \overleftarrow\exp{\left[\, \int^{\xi_1}_{\xi_2} %
\widehat\Lambda({\bf A}^*(\xi),{\bf A}(\xi))\, %
d\xi\right]}\,\right\rangle_0 \, \Phi_{\beta}^{\prime}(\rho) %
\,c(x) \,\,
\end{array}
\end{equation}
At that, according to definitions of %
holomorphic path integration and our related  %
seed averaging \,$\langle\dots\rangle_0$\, (\ref{cfa}), %
we can turn back to the pseudo-quantum and then %
original description of BP's motion and rewrite %
expression (\ref{rep}) as
\begin{equation}
\begin{array}{l}
\mathcal{G}_{\alpha\beta}(\tau)\,=\,  \label{rep1}
\left\langle 0\left|\, %
\frac {u_0^2}{T^2}\int c(x)\,\Phi^{\prime}_{\alpha}(\rho)\, %
\, \exp{\,[\,\tau\,  %
\widehat\Lambda({\bf A}^\dag,{\bf A})\,]}\, %
\, \Phi_{\beta}^{\prime}(\rho) %
\,c(x)\,dx\,\, \right|0\right\rangle \,=\, \\
\,=\, \int d{\bf P} \,\, \breve{\mathcal{G}}_{\alpha\beta} %
(\tau,\, V,\nabla_P)\,  G_M({\bf P})\,\, \,
\end{array}
\end{equation}
In the latter row we introduced operators
\begin{equation}
\begin{array}{l}
\breve{\mathcal{G}}_{\alpha\beta}(\tau,\, V,\nabla_P)\,\equiv\, %
\frac {\nu}{TM} \int  %
\,\Phi^{\prime}_\alpha(\rho)\,\, %
\exp{[\,\tau\, \breve\Lambda(V,\nabla_P)\,]}\, %
\, \Phi^{\prime}_\beta(\rho)\,g(x)\,dx\,\,, \, \label{rep2}\\
\breve\Lambda(V,\nabla_P)\,=\, %
({\bf V} -{\bf v})\cdot\nabla\,+\, %
\Phi^{\prime}(\rho)\cdot(\nabla_p -\nabla_P)\,\,, \, %
\end{array}
\end{equation}
with designations\,  %
\,$\,g(x) = G_m({\bf p})\,E(\rho)\,$,\, %
\,$\nabla =\partial/\partial \rho $\,,\, %
\,$\nabla_p =\partial/\partial {\bf p} $\,,\, %
\,$\nabla_P =\partial/\partial {\bf P} $\,\, %
(we also used definitions of BP's %
boson operators (\ref{aa}) and its ``vacuum'':\,   %
\,$|0\rangle = G_M({\bf P})$\,,\, %
\,$\langle 0| = \int d{\bf P}$\,\,).

Combining this kernel's approximation with exact %
Eq.\ref{sv}, we come to Gaussian path integral %
and, - like in Sec.6 of \cite{p0806}, -  obtain
\begin{equation}
\begin{array}{l}
V_0(t,i{\bf k};\nu,f)\,\approx\, %
\exp\,\left[\,u_0^2 \int_{t>\,\xi_1>}\int_{\xi_2>\,0} %
i{\bf k}\cdot %
Q(\xi_1-\xi_2)\cdot %
\left(i{\bf k} + \frac {{\bf f}}T \right) %
\,d\xi_2\,d\xi_1\,\right]\, \, \label{asv}\\
\end{array}
\end{equation}
with kernel \,$Q(\xi)\,$ determined by formulae
 \begin{equation}
\begin{array}{l}
Q\,=\,\Theta\, - \,\Theta\otimes\mathcal{G}\otimes\Theta\, +\, %
\Theta\otimes\mathcal{G}\otimes\Theta\otimes %
\mathcal{G}\otimes\Theta \,-\, \dots\,\,, \,\,\\
\int_0^\infty e^{-z\tau}\,Q(\tau)\,d\tau\,=\, %
\left[\,z\,+\,\int_0^\infty e^{-z\tau}\,\mathcal{G}(\tau)\,d\tau\, %
\right]^{-\,1}\,\,,\,\label{conv}
\end{array}
\end{equation}
and\, $\,\otimes\,$ symbolizing causal time convolution. %

\,\,\,

{\bf 3}.\,\,
Obviously, first, both the matrix functions\, %
$\,\mathcal{G}(\tau)\,$ and $\,Q(\tau)\,$\, are diagonal: %
$\,\mathcal{G}_{\alpha\beta}= \delta_{\alpha\beta}\mathcal{G}\,$,\, %
$\,Q_{\alpha\beta}= \delta_{\alpha\beta}Q\,$. %
Second, under our above assumptions about the %
interaction potential, $\,\mathcal{G}(\tau)\,$ is fast %
decaying function of \,$\tau$\,, vanishing at %
\,$\tau\gtrsim \tau_c\sim r_0/v_0^\prime$\,,\, where %
\,$\tau_c$\, is typical collision duration, %
\,$r_0$\, is characteristic radius of BP-atom interaction, %
and \,$v_0^\prime =\sqrt{T/m^\prime}$\, with %
BP-atom reduced mass\, \,$m^\prime =mM/(m+M)$\,.

Hence, if being interested in BP's observation times %
much longer than mean BP's velocity relaxation time, %
or BP's mean free path time, \,$\tau_0$\,, we can make %
replacements
\begin{equation}
\begin{array}{l}
\mathcal{G}(\tau)\,\Rightarrow\, %
\overline\gamma\,\delta_+(\tau)\,\,, \,\,\,\,\,\,
\int_0^\infty e^{-z\tau}\,\mathcal{G}(\tau)\,d\tau\, %
\Rightarrow\, \overline\gamma\,\,, \label{lt}
\,\,\,\,\,\, \int_0^\infty e^{-z\tau}\,Q(\tau)\,d\tau\,\Rightarrow\, %
\frac 1{z+\overline\gamma}\,\,,\,
\\  
\overline\gamma\,= \int_0^\infty \mathcal{G}(\tau)\, d\tau\, %
=\,\frac {2m}{M+m}\,\,\nu\,\int\int |{\bf v} -
{\bf V}|\,\Sigma(|{\bf v} - {\bf V}|)\, %
\,G_m({\bf p})\,G_M({\bf P})\,d{\bf p}\, %
d{\bf P}\, 
\end{array}
\end{equation}
Here \,$\Sigma$\, is effective full relative %
velocity-dependent cross-section of BP-atom collision,\, %
\,$\Sigma\sim \pi r_0^2$\,. Evidently, %
\,$\overline\gamma$\, is nothing %
but mean BP's velocity relaxation rate. %
Therefore we may write \,$\overline\gamma=1/\tau_0$\, and %
say also that \,$\overline\gamma$\, is characteristic ``probability %
of collision per unit time''.

These simplifications imply
\begin{equation}
\begin{array}{l}
V_0(t,i{\bf k};\nu,f)\,\approx\, %
\exp\,\left[\,u_0^2 \int_{t>\xi_1>\xi_2>0} %
i{\bf k}\,\,e^{-\overline\gamma\,(\xi_1-\xi_2)}\, %
\left(i{\bf k} + \frac {{\bf f}}T \right) %
\right]\,\,\approx\, %
e^{\,i{\bf k}\,D_0\,(i{\bf k} + %
{\bf f}/T)\,t}\,\,, \, \label{fra}
\end{array}
\end{equation}
where\, \,$D_0=u_0^2/\overline{\gamma} =u_0^2\tau_0$\, is %
BP's diffusivity, and the latter exponential corresponds to  %
time-independent wave (probe) vector and time-independent force.

\section{Improved consideration and %
time-scaleless relaxation rate fluctuations}

{\bf 1}.\,\,
The result (\ref{fra}) of above considered approximation (\ref{rep}) %
is standard diffusive random process (``Ornstein-Uhlenbeck %
process'') possessing purely Gaussian statistics with %
exponential velocity correlation function.
Thus, we neglected possibly non-exponential %
behavior of this function, that is non-linearity of BP's velocity %
relaxation. However, from viewpoint of diffusion law, - %
i.e. BP's total path probability distribution %
at \,$t\gg \tau_0$\,, - %
this non-linearity is quite insignificant since %
does not cancel  %
the Gaussian long-range asymptotic of %
\,$V_0(t,\Delta R;\nu,f)$\,.

Of course, the non-linearity may induce some dependence %
of BP's diffusivity and mobility on the force \,$f$\,. %
At that, nevertheless,diffusivity and mobility remain well %
certain quantities, one  and the same for all %
possible realizations of BP's random walk and all %
(long enough) fragments of any particular realization.

It would be quite another matter if beyond the %
considered approximation (\ref{rep}) we found %
qualitatively different random walk which has  %
no certain diffusivity and mobility. %
Below, we want to argue that just such picture arises %
from analysis of third of the three above mentioned %
factors, that is interference of BP's collisions. %

\,\,\,

{\bf 2}.\,\,
First, it is useful to see mere existence of %
``interference of collisions'' or, %
in other words, essential statistical correlations %
between them.

With this purpose, consider pair self-correlation %
of the kernel in Eq.\ref{sv},
\begin{equation}
\begin{array}{l}
\mathcal{G}^{(2)}(\xi_1,\xi_2;\xi_3,\xi_4)\,\equiv\, %
\langle\, %
\mathbb{G}\{\xi_1,\xi_2,\,{\bf A}^*,{\bf A}\}\, %
\mathbb{G}\{\xi_3,\xi_4,\,{\bf A}^*,{\bf A}\}\, %
\rangle_0\,\,, \label{pc}\,
\end{array}
\end{equation}
omitting indices for brevity, and taking in mind  %
most interesting variant of time-separated %
intervals \,$(\xi_1,\xi_2)$\, and \,$(\xi_3,\xi_4)$\,, %
e.g. when \,$\xi_1>\xi_2>\xi_3>\xi_4$\,. %
Then, clearly, the function %
\,$\mathcal{G}^{(2)}(\xi_1,\xi_2;\xi_3,\xi_4)$\, %
delegates mutual correlation of two different collisions.

If this function reduced to product of the averaged kernels, %
\,$\mathcal{G}(\xi_1-\xi_2)$\, and %
\,$\mathcal{G}(\xi_3-\xi_4)$\,, at any or some finite %
(of order of \,$\tau_0$\,) \,$\xi_2-\xi_3$\,, then %
it would be substantial ground for the above %
``fixed relaxation rate approximation''. %
In fact, however, %
\,$\mathcal{G}^{(2)}(\xi_1,\xi_2;\xi_3,\xi_4)$\, %
has no natural tendency to such reduction.
Indeed, let us write out the average (\ref{pc}) in more detail %
in the form
\begin{equation}
\begin{array}{l}
\mathcal{G}^{(2)}(\xi_1,\xi_2;\xi_3,\xi_4)\,=\, %
\left(\frac {u_0^2}{T^2}\right)^2 %
\int\!\!\int c(x_1)\,c(x_2)\,\Phi^{\prime}(\rho_1)\, %
\Phi^{\prime}(\rho_2)\, %
\,\langle\,\,
\overleftarrow\exp{[\, \int^{\xi_1}_{\xi_2} %
\widehat\Lambda_1({\bf A}^*(\xi),{\bf A}(\xi))\, %
d\xi\,]}\,\, %
\times \,\label{g2}\\ \times\,\, %
\overleftarrow\exp{[\, \int^{\xi_3}_{\xi_4} %
\widehat\Lambda_2({\bf A}^*(\xi),{\bf A}(\xi))\, %
d\xi\,]}\, \,\rangle_0\,\, %
\Phi^{\prime}(\rho_1)\,\Phi^{\prime}(\rho_2)\, %
c(x_1)\,c(x_2) \,\,, \,
\end{array}
\end{equation}
where subscripts of operators %
\,$\widehat\Lambda_{1,2}$\, indicate that they act %
in different spaces \,$x_1$\, and \,$x_2$\,.
We can unify these two operators into single %
but additionally time-dependent one, %
\,$\widehat\Lambda (\xi;{\bf A}^*,{\bf A})$\,, %
which coincides with \,$\widehat\Lambda_{1,2}$\, in %
intervals \,$(\xi_1,\xi_2)$\, and \,$(\xi_3,\xi_4)$\,, %
respectively, and turns to zero otherwise. %
Then (\ref{g2}) transforms to
\begin{equation}
\begin{array}{l}
\mathcal{G}^{(2)}(\xi_1,\xi_2;\xi_3,\xi_4)\,=\, %
\left(\frac {u_0^2}{T^2}\right)^2 %
\int\!\!\int c(x_1)\,c(x_2)\,\Phi^{\prime}(\rho_1)\, %
\Phi^{\prime}(\rho_2)\,\, %
\times \,\label{g2_}\\
\times\,\, \left\langle\,
\overleftarrow\exp{[\, \int %
\widehat\Lambda(\xi;{\bf A}^*(\xi),{\bf A}(\xi))\, %
d\xi\,]}\,\right\rangle_0\, %
\Phi^{\prime}(\rho_1)\,\Phi^{\prime}(\rho_2)\, %
c(x_1)\,c(x_2) \,\,\,
\end{array}
\end{equation}
Here again, similarly to (\ref{rep})-(\ref{rep1}), %
we can make transition back to pseudo-quantum %
representation and rewrite expression (\ref{g2_}) first as
\begin{equation}
\begin{array}{l}
\mathcal{G}^{(2)}(\xi_1,\xi_2;\xi_3,\xi_4)\,=\, %
\left(\frac {u_0^2}{T^2}\right)^2 %
\int\!\!\int c(x_1)\,c(x_2)\,\Phi^{\prime}(\rho_1)\, %
\Phi^{\prime}(\rho_2)\, %
\,\times\,\label{g22}\\ \,\times\,\, %
\left\langle 0\left|\,\,  %
\overleftarrow\exp{[\, \int %
\widehat\Lambda(\xi;{\bf A}^\dag,{\bf A})\, %
d\xi\,]}\,\, \right|0\right\rangle\,\, %
\Phi^{\prime}(\rho_1)\,\Phi^{\prime}(\rho_2)\, %
c(x_1)\,c(x_2) \,\,\,
\end{array}
\end{equation}
and then, obviously, as
%
\begin{equation}
\begin{array}{l}
\mathcal{G}^{(2)}(\xi_1,\xi_2;\xi_3,\xi_4)\,=\, %
\mathcal{G}^{(2)}(\xi_1-\xi_2\,;\,\xi_3-\xi_4)\,\equiv\, %
\,\langle 0|\, \widehat{\mathcal{G}}(\xi_1-\xi_2)\, %
\widehat{\mathcal{G}}(\xi_3-\xi_4)\, %
|0\rangle\, \,, \,\label{g2o}
\end{array}
\end{equation}
with \,$\widehat{\mathcal{G}}$\, being quantum operator %
analogue of the kernel \,$\mathcal{G}$\,, namely,
\begin{equation}
\begin{array}{l}
\widehat{\mathcal{G}}_{\alpha\beta}(\tau)\,\equiv\, %
\widehat{\mathcal{G}}_{\alpha\beta}(\tau,\, %
{\bf A}^\dag,{\bf A})\,\equiv\, %
\frac {u_0^2}{T^2}\int c(x)\,\Phi^{\prime}_{\alpha}(\rho)\, %
\, \exp{\,[\,\tau\,  %
\widehat\Lambda({\bf A}^\dag,{\bf A})\,]}\, %
\, \Phi_{\beta}^{\prime}(\rho) %
\,c(x)\, dx \,\, \label{ok}
\end{array}
\end{equation}
Or, equivalently, returning to our basic notations,
\begin{equation}
\begin{array}{l}
\mathcal{G}^{(2)}(\tau;\,\tau^\prime)\,=\, %
\int d{\bf P}\,\,  %
\breve{\mathcal{G}}(\tau, V,\nabla_P)\, %
\breve{\mathcal{G}}(\tau^\prime, V,\nabla_P)\, %
\,G_M({\bf P})\,\, \label{g2o_}
\end{array}
\end{equation}
with operator \,$\breve{\mathcal{G}}$\, already defined %
in (\ref{rep1})-(\ref{rep2}). Clearly, %
according to (\ref{aa}),
\begin{equation}
\begin{array}{l}
\breve{\mathcal{G}}(\tau,\, V,\,\nabla_P)\,=\, %
\widehat{\mathcal{G}}(\tau,\, -\sqrt{TM}\,\nabla_P,\, %
\sqrt{M/T}\,(V+T\nabla_P))\, \,\, \label{gg}
\end{array}
\end{equation}

From Eq.\ref{g2o} it visually follows that %
the correlation function of interaction kernel %
is indifferent to time separation of the %
intervals \,$(\xi_1,\xi_2)$\, and \,$(\xi_3,\xi_4)$\, %
where collisions take place. Clearly, if we considered %
third- and higher-order statistical moments of %
the kernel \,$\mathbb{G}$\,, that is
\[
\mathcal{G}^{(n)}(\xi_1,\xi_2;\,\dots\, ;\xi_{2n-1},\xi_{2n})\, %
\equiv\, \langle\, \mathbb{G}\{\xi_1,\xi_2,A^*,A\}\, %
\dots\, \mathbb{G}\{\xi_{2n-1},\xi_{2n},A^*,A\}\, %
\rangle_0\,\,,\, 
\]
we quite similarly %
would find their independence on time distances %
between collision intervals:
\begin{equation}
\begin{array}{l}
\mathcal{G}^{(n)}(\xi_1,\xi_2;\,\dots\, ;\xi_{2n-1},\xi_{2n})\,=\, %
\mathcal{G}^{(n)}(\xi_1-\xi_2;\,\dots\, ;\xi_{2n-1}-\xi_{2n})\,  %
\equiv\, \\ \,\equiv\, %
\,\langle 0|\, \widehat{\mathcal{G}}(\xi_1-\xi_2)\, %
\dots\, \widehat{\mathcal{G}}(\xi_{2n-1}-\xi_{2n})\, %
|0\rangle\, =\, \,\,\label{gn}\\
\,=\, \int d{\bf P}\,\,  %
\breve{\mathcal{G}}(\xi_1-\xi_2, V,\nabla_P)\, %
\dots\, \breve{\mathcal{G}}(\xi_{2n-1}-\xi_{2n}, V,\nabla_P)\, %
\,G_M({\bf P})\,\,
\end{array}
\end{equation}
for mutually non-intersecting intervals %
\,$(\xi_{2j-1},\xi_{2j})$\,.

Thus, BP's relaxation rate %
undergoes non-decaying, or ``infinitely long-living'', %
fluctuations. Notice that this could   %
be predicted merely as manifestation of non-decaying %
character of the ``seed'' correlators (\ref{sm}).

\,\,\,

{\bf 3}.\,\,
Generally, existence of non-decaying relaxation rate (RR) %
fluctuations does not prevent diffusive chaotic %
behavior of BP's walk, since collisions  %
make their work at any values of \,$\mathcal{G}$\,   %
in (\ref{conv}) comparable with \,$\overline\gamma$\, from (\ref{lt})).
Non-decaying, - or time-scaleless, - RR fluctuations just  %
make BP's walk purely chaotic:\, %
it is becomes so much random that even has no certain %
({\it \,a priori\,} predictable) ``probability of %
collisions per unit time''.

Such free behavior of actual (physical) %
random walks is origin of 1/f\,-type fluctuations %
in their diffusivity and mobility, %
which can be explained and described  %
\cite{pr157,bk1,bk2,ufn,pr195} also as logically inevitable %
property of such many-particle dynamical systems %
constantly forgetting their past (e.g. %
forgetting pre-history of collisions or other kinetic and %
transport events). Thus, relaxation always %
appears together with its scaleless fluctuations %
\cite{i1,i2,i3,p1,p0802,tmf,kmg,e2,p1207,p1302,ufn1}.

In essence, such logics was suggested already by N.\,Krylov %
\cite{kr} who visually connected phenomenon of  relation to  %
exponential instability and ``mixing'' of system's phase trajectories %
and simultaneously pointed out that %
these features in no way presume appearance of some  %
{\it \,a priori\,} definite (phase trajectory-independent) %
relaxation rates (relative frequencies of kinetic events) %
and ``probabilistic laws''\, %
\footnote{\, %
Such the logics is exact antithesis of the %
classical Bernoulli's one \cite{jb} based on conjecture %
that seemingly (physically) independent events are %
are statistically independent.
In essence, it is nothing but %
implicit assumption that relative %
frequency of (one or another sort of) %
events always has definite limit under constant %
conditions and large enough number of observations (trials). %

Although in general this assumption is logically incompatible %
with statistical mechanics \cite{kr}, %
the Bernoulli's logics %
still dominates over scientific thinking  %
(so that, for example, well known monograph \cite{rl}, %
devoted to dynamical grounds of kinetics, %
starts from vocabulary of stochastic processes).
}\,. %

In more concrete terms, %
a small change of initial BP's velocity \,$V(0)$\, %
leads to exponentially growing %
changes of its velocity \,$V(t)$\, and, all the more, %
its path \,$\Delta R(t)$\,  %
after a sequence of BP's collisions %
with atoms. At that, characteristic relaxation times %
\,$\tau_+=\partial \Delta R(t)/\partial V(0)$\,   %
and \,$\tau_-=-\partial \Delta R(t)/\partial V(t)$\,  %
are essentially random quantities, - such that %
\,$\langle\tau_+^2\rangle/\tau_0^2 = %
\langle\tau_-^2\rangle/\tau_0^2  %
\rightarrow\infty\,$ nearly exponentially at %
\,$t/\tau_0\rightarrow\infty$\,, - and essentially %
correlated one with another, so that\,  %
\,$\langle\tau_+\tau_-\rangle - %
\langle\tau_+\rangle \langle\tau_-\rangle %
\sim\tau_0^2$\,\, regardless of how long is the observation time
\footnote{\, %
More precisely, this cross-correlation %
may decrease or increase with time by a slow (logarithmic %
or power) law.
}\,.
This is nearly what is said by Eqs.\ref{g2o} and \ref{gn}
\footnote{\, %
In abstract words, forgetting the past %
in the course of relaxation means %
uncontrollability of its rate and absence of a %
limiting upper time scale in the rate fluctuations %
(in above consideration, this is reflected by indifference of %
the kernel \,$\mathbb{G}$\,'s correlators to time %
distances \,$\xi_{2j}-\xi_{2j+1}$\,).
}\,. %

\section{Free relaxation rate approximation}

{\bf 1}.\,\, %
Let us leave ``philosophy'' and %
return to the exact expression (\ref{sv}). %
We can rewrite it in the form
\begin{equation}
\begin{array}{l}
V_0(t,i{\bf k};\nu,f)\,=\,\langle\, %
\exp\,\{\int^t_0 %
u_0\,[\,i{\bf k}\cdot {\bf A}(\xi)+ %
(i{\bf k} + {\bf f}/T ) %
\cdot {\bf A}^*(\xi)\, ]\,d\xi\, %
-\, \\ \,\, - %
\int_{t>\,\xi_1>}\int_{\xi_2>\,0} {\bf A}^*(\xi_1)
\cdot\mathbb{G}\{\xi_1,\xi_2,{\bf B}^*,{\bf B}\} %
\cdot {\bf A}(\xi_2) \,d\xi_2\,d\xi_1 %
\,\}\,\rangle_0\,\,  \label{svn}
\end{array}
\end{equation}
with new additional %
random variables \,$\{{\bf B}^*,{\bf B}\}$\, in place of %
\,$\{{\bf A}^*,{\bf A}\,$\, inside the kernel, if these %
new variables are statistically identical to the %
old ones. This means that
\begin{equation}
\begin{array}{l}
\langle\,{\bf B}_{\alpha}(t_1)\,{\bf B}_{\beta}(t_2)\, %
\rangle_0\,= \,\langle\,{\bf B}^*_{\alpha}(t_1)\,{\bf
B}^*_{\beta}(t_2)\, \rangle_0\,=\,0\,\,\,, \,\,\,\,\,\,\langle\,{\bf
B}_{\alpha}(t_1)\, {\bf B}^*_{\beta}(t_2)\,\rangle_0\,
=\,\delta_{\alpha\beta}\,\Theta(t_1-t_2)\,\,\,,\label{smb}
\end{array}
\end{equation}
and, clearly,
\begin{equation}
\begin{array}{l}
\langle\,{\bf A}_{\alpha}(t_1)\,{\bf B}_{\beta}(t_2)\, %
\rangle_0\,= %
\,\langle\,{\bf A}^*_{\alpha}(t_1)\,{\bf B}^*_{\beta}(t_2)\,
\rangle_0\,=\,0\,\,\,, \,\\
\langle\,{\bf A}_{\alpha}(t_1)\,
{\bf B}^*_{\beta}(t_2)\,\rangle_0\,=\, %
\langle\,{\bf B}_{\alpha}(t_1)\,
{\bf A}^*_{\beta}(t_2)\,\rangle_0\, %
=\,\delta_{\alpha\beta}\,\Theta(t_1-t_2)\,\,\, \label{smab}
\end{array}
\end{equation}

Then, taking in mind our previous reasonings, %
we may say that\, (i) \,${\bf A}\,-\,{\bf A}^*$\,'s inter-correlations %
in Eq.\ref{svn} are responsible for relaxation of %
BP's velocity and resulting diffusive BP's motion %
instead of ballistic one,\, (ii) %
\,${\bf B}\,-\,{\bf B}^*$\,'s inter-correlations %
represent randomness of the relaxation rate (RR) in itself, - %
that is ``interference of collisions'', - %
while\, (iii) cross-correlations \,${\bf A}\,-\,{\bf B}^*$\,  %
and \,${\bf B}\,-\,{\bf A}^*$\, reflect statistical influence  %
of BP's walk onto RR and its fluctuations. %

The latter item includes %
the second of three above mentioned factors. %
It is not critical for BP's walk, at least  %
in qualitative sense. Therefore we can %
consider approximation which excludes this factor by neglecting the %
\,${\bf A}\,-\,{\bf B}^*$ and ${\bf B}\,-\,{\bf A}^*$\, %
cross-correlations, i.e. turning all right-hand sides in %
(\ref{smab}) to zero.
This formal simplification again, - similar to the ``fixed RR %
approximation'' in Sec.4, - makes BP's %
velocity relaxation forcedly linear. %
However, now relaxation rate (RR) and hence %
BP's diffusivity are not fixed but random quantities, %
although they will be treated with more or %
less significant losses because of the same simplification.
In particular, loss of fine temporal structure of %
BP's random walk and thus excess idealization %
of concept of RR fluctuations, as if RR could be measured %
irrespectively of the walk. In this sense, the resulting model  %
can be named ``free RR approximation''.

\,\,\,

{\bf 2}.\,\,
Accepting such approximation, clearly, we can %
perform in Eq.\ref{svn} averaging first in respect to %
\,$\{{\bf A}^*,\,{\bf A}\}$\, and later in respect to %
\,$\{{\bf B}^*,\,{\bf B}\}$\,. The first operation yields
\begin{equation}
\begin{array}{l}
V_0(t)\,\approx\, \left\langle\, %
\exp\,\left[\,u_0^2 \int_{t>\,\xi_1>}\int_{\xi_2>\,0} %
i{\bf k}\cdot  \mathbb{Q}\{\xi_1,\xi_2,\, %
{\bf B}^*,{\bf B}\}\cdot %
\left(i{\bf k} + \frac {{\bf f}}T \right) %
\,d\xi_2\,d\xi_1\,\right]\, %
\right\rangle_0\, \, \label{asvn}
\end{array}
\end{equation}
with kernel \,$\mathbb{Q}\{\xi_1,\xi_2, %
{\bf B}^*,{\bf B}\}\,$ now given by %
infinite series of convolutions,
\begin{equation}
\begin{array}{l}
\mathbb{Q}\,=\,\Theta\, - \,\Theta\otimes\mathbb{G}\otimes\Theta\, +\, %
\Theta\otimes\mathbb{G}\otimes\Theta\otimes %
\mathbb{G}\otimes\Theta \,-\, \dots\,\,, \,\label{convn}
\end{array}
\end{equation}
which generalizes series from (\ref{conv}) %
to functional \,$\mathbb{Q}=\mathbb{Q}\{\xi_1,\xi_2, %
{\bf B}^*,{\bf B}\}\,$ in place of the function %
\,$\mathcal{Q}(\xi_1-\xi_2)$\,. Equivalently,
\begin{equation}
\begin{array}{l}
\mathbb{Q}\{t_1,t_2,\,B^*,B\}=\,\Theta(t_1-t_2) - %
\!\!\int_{t_1>\xi_1>\xi_2>t_2}\!\! \Theta(t_1-\xi_1)\, %
\mathbb{G}\{\xi_1,\xi_2,\,B^*,B\}\, %
\mathbb{Q}\{\xi_2,t_2,\,B^*,B\}\, %
d\xi_2\,d\xi_1\,\, \,\,\,\label{convn_}
\end{array}
\end{equation}

The exponential in Eq.\ref{asvn} represents Gaussian %
random walk with randomly varying diffusivity and mobility %
(diffusion and drift rates).
Importantly, that are long-living, or %
time-scaleless, variations, which follows from %
previous section and merely from %
non-decaying of the \,$B$ - $B^*$\,  correlator (\ref{smb}).

Unfortunately, result of averaging of %
the exponential in Eq.\ref{asvn} %
hardly can be written in transparent or at least %
generally closed form. %
Hence, we need in further simplifications. %
Next, let us consider some possibilities.

\section{Free RR approximation for low-density gas}

{\bf 1}.\,\, %
Naturally, main difficulties in calculation of %
the average (\ref{asvn}) %
come from time non-locality of the keenel %
\,$\mathbb{G}\{\xi_1,\xi_2,{\bf B}^*,{\bf B}\}$\,.
They become lighten, however, if gas is rare %
enough, that is \,$\nu r_0^3\ll 1$\, and hence  %
\,$\tau_c/\tau_0\ll 1$\, (all the more in the  %
low-density, or Boltzmann-Grad, limit). %
Then, apparently, we can try to neglect %
contributions of mutual time intersections %
of factors  \,$\mathbb{G}$\, in power series expansion %
of the exponential in (\ref{asvn}).

At that, \,$\mathbb{G}\{\xi_1,\xi_2,B^*,B\}$\, %
is treated as delta-function of %
\,$\xi_1-\xi_2$\,, that is, similar to (\ref{lt}),
\begin{equation}
\begin{array}{l}
\mathbb{G}\{t^\prime,t^{\prime \prime},\,B^*,B\}\, \Rightarrow\, %
\Gamma\{t^\prime,\,B^*,B\}\, %
\delta_+(t^\prime-t^{\prime \prime})\,\,, \, \label{bg}
\end{array}
\end{equation}
where, nevertheless, \,$\Gamma\{t^\prime,\,B^*,B\}$\, is thought %
as definite functional of \,$\{B^*(\xi),\,B(\xi)\}$\, %
with \,$\xi$\, from some time interval \,$\sim \tau_c$\, %
around \,$t^\prime$\,.
Moreover, when total observation time is much greater than %
\,$\tau_0$\,, then the kernel %
\,$\mathbb{Q}\{\xi_1,\xi_2,B^*,B\}$\, %
also may be treated in time-local fashion, with replacement
\begin{equation}
\begin{array}{l}
u_0^2\,\mathbb{Q}\{t^\prime,t^{\prime \prime},\, %
B^*,B\}\, \Rightarrow\, %
\mathbb{D}\{t^\prime,\,B^*,B\}\, %
\delta_+(t^\prime-t^{\prime \prime})\,\,, \, \label{bq}
\end{array}
\end{equation}
where \,$\mathbb{D}\{t,\,B^*,B\}$\, is functional depending  %
mainly on \,$\{B^*(\xi),\,B(\xi)\}$\, %
with \,$|\xi -t |\sim \tau_0$\, %
and representing random (fluctuating) BP's diffusivity.

\,\,\,

{\bf 2}.\,\, %
The dilute gas approximation (\ref{bg}) allows us to use relations %
(\ref{rep1}),(\ref{g2o}) and (\ref{gn}) and  %
replace the averaging in (\ref{asvn}) by  %
equivalent operations in the original %
or pseudo-quantum formalisms.
In the original form, we have first to write out operator %
analogue of Eq.\ref{bg}:
\begin{equation}
\begin{array}{l}
\breve{\mathcal{G}}(\tau,\, V,\nabla_P) %
\, \Rightarrow\, %
\breve\Gamma(V,\nabla_P)\, %
\delta_+\tau)\,\,, \, \label{bgo}\,\,\,\,\,\,
\breve\Gamma\,=\, \int_0^\infty \breve{\mathcal{G}}(\tau)\, %
d\tau\,\,, \,
\end{array}
\end{equation}
with operator \,$\breve{\mathcal{G}}$\, %
defined in (\ref{rep2}), %
and \,$\breve\Gamma(V,\nabla_P)\,$ playing role of %
RR operator.

Combining formulae (\ref{rep1}),(\ref{g2o}) and (\ref{gn}) %
with (\ref{bgo}) and substituting them to Eq.\ref{asvn}, %
we arrive to
\begin{equation}
\begin{array}{l}
V_0(t)\,\approx\, \int d{\bf P}\,\, %
\exp\,\left[\,u_0^2 \int_{t>\,\xi_1>}\int_{\xi_2>\,0} %
i{\bf k}\cdot  \breve{\mathcal{Q}}(\xi_1-\xi_2,\, %
{\bf V},\nabla_P)\cdot %
\left(i{\bf k} + \frac {{\bf f}}T \right) %
\,d\xi_2\,d\xi_1\,\right]\,\, %
G_M({\bf P})\, \, \label{asvo}
\end{array}
\end{equation}
The operator kernel \,$\breve{\mathcal{Q}}$\, here  %
is directly determined by Eqs.\ref{convn}-\ref{convn_}   %
together with (\ref{gn}),
\begin{equation}
\begin{array}{l}
\breve{\mathcal{Q}}(\tau,\,V,\nabla_P)\,=\, %
e^{-\,\breve{\Gamma}(V,\nabla_P)\,\tau}\,\, \label{bqo}
\end{array}
\end{equation}
At constant wave vector and force and at %
sufficiently large observation time, %
expression (\ref{asvo}) turns to
\begin{equation}
\begin{array}{l}
V_0(t,ik;\nu,f)\,\approx\, \int d{\bf P}\,\, %
e^{\,i{\bf k}\cdot  \breve{\mathcal{D}} %
({\bf V},\nabla_P)\cdot %
(i{\bf k} + {\bf f}/T)\, t}\,\, %
G_M({\bf P})\,\,,  \, \label{do}\\
\breve{\mathcal{D}}({\bf V},\nabla_P)\,=\,u_0^2\, %
\breve{\Gamma}^{-1}({\bf V},\nabla_P)\,\,,
\end{array}
\end{equation}
with \,$\breve{\mathcal{D}}({\bf V},\nabla_P)$\, in the %
role of BP's diffusivity operator %
(operator image of \,$\mathbb{D}\{t^\prime,B^*,B\}$\, %
from Eq.\ref{bq}). %

To go to the pseudo-quantum representation, %
we have to use relation (\ref{gg}) and replacements
\begin{equation}
\begin{array}{l}
\breve\Gamma(V,\nabla_P)\, \Leftrightarrow\, %
\widehat\Gamma(A^\dag,A)\,=\, %
\int_0^\infty \widehat{\mathcal{G}}(\tau)\,d\tau\, %
\,, \\  %
\breve{\mathcal{Q}}(\tau,V,\nabla_P)\,, \, %
\breve{\mathcal{D}}(V,\nabla_P)\,\, %
 \Leftrightarrow\,\, %
\widehat{\mathcal{Q}}(\tau,A^\dag,A)\,=\, %
\exp{[\,-\tau\, \widehat\Gamma(A^\dag,A)]}\,\,, \,\, %
\widehat{\mathcal{D}}(A^\dag,A)\,=\, %
u_0^2\,[\,\widehat\Gamma(A^\dag,A)\,]^{-1}\,\,, \, \label{eqv}\\
\int d{\bf P}\,\dots\, G_M({\bf P})\,\,\,  %
\,\, \,\Leftrightarrow\,\,\, %
\langle 0|\, \dots\, |0\rangle\,\,  \,\,\,
\end{array}
\end{equation}


\,\,\,

{\bf 3}.\,\,
It is necessary to notice that, strictly speaking, %
just made application of formulae %
(\ref{rep1}),(\ref{g2o}) and (\ref{gn}) %
requires chronological time ordering of the operator %
images \,$\breve\Gamma(V,\nabla_P)$\, of functionals %
\,$\Gamma\{t,\,B^*,B\}$\,.
Although these operators %
get no literal time dependence, such the requirement %
creates obstacles to ``disentangling''  of %
operator images of time-intersecting copies of %
\,$\mathbb{G}\{t^\prime,t^{\prime \prime},\,B^*,B\}$\,.  %
The matter is that  %
\,$\breve\Gamma(V,\nabla_P)$\, along with  %
\,$\Gamma\{t,\,B^*,B\}$\, is tensor-like (matrix-like) %
object whose components may be mutually %
non-commuting.

By these reason, transition %
from Eq.\ref{asvn} to Eq.\ref{asvo} %
is approximate, - together with (\ref{bq}), %
(\ref{do}) and other related formulas, - %
even under the dilute gas condition.


\,\,\,

{\bf 4}.\,\,
In view of this formal difficulty, %
it seems reasonable to supplement the ``free RR'' %
and ``dilute gas'' approximations  %
with one more simplification. Namely, %
replacing the originally multi-component matrix RR operator %
by its scalar diagonal part:
\begin{equation}
\begin{array}{l}
\breve\Gamma_{\alpha\beta}\,\Rightarrow\, %
\delta_{\alpha\beta}\,\breve\Gamma\,\,, \,\label{diag} %
\,\,\,\,\,  %
\breve\Gamma\,\Rightarrow\, %
\frac 13 \sum_{\alpha}\int_0^\infty %
\breve{\mathcal{G}}_{\alpha\alpha}(\tau)\,d\tau\, \,\,
\end{array}
\end{equation}
After that the difficulty disappears. %

We can justify this simplification by the fact that %
due to the gas isotropy the RR operator %
is natively isotropic, that is obeys spherical symmetry.
Therefore, replacement (\ref{diag}) itself %
must not cause too big qualitative losses.
Then, combining (\ref{do}) and (\ref{diag}) and %
performing in (\ref{do}) Fourier transform %
over \,${\bf k}$\,, we obtain the %
diffusion law in the form
\begin{equation}
\begin{array}{l}
V_0(t,\Delta R;\nu,f)\,\approx\,
\int d{\bf P}\,\, %
\left(\frac {\breve\Gamma} %
{4\pi u_0^2 t}\right)^{3/2}\, %
\exp{\left[\,- \frac {\breve\Gamma}{4\, u_0^2 t}\, %
\left(\Delta R - u_0^2\,\breve\Gamma^{-1}\, %
\frac fT \right)^2\,\right]}\,\, %
G_M({\bf P})\,\,,  \, \label{dl}
\end{array}
\end{equation}
with scalar RR operator, in the sense of (\ref{diag}). %
Its properties, however, will be discussed in parallel %
with that of the full matrix RR operator.


\section{Relaxation rate operator, %
relaxation rate distribution, and diffusion law}

{\bf 1}.\,\,
From viewpoint of Eqs.\ref{bgo}-\ref{do} and \ref{dl}, %
spectral properties of the RR operator %
\,$\breve\Gamma$\, are of most importance. %
To consider them, notice, first, that the Liouville operator %
\,$\breve\Lambda(V,\nabla_P)$\, commutes with  %
multiplication by \,$g(x)\,G_M(P)$\,, i.e. %
\,$\breve\Lambda\,g(x)\,G_M(P)\,-\, %
g(x)\,G_M(P)\,\breve\Lambda\, =0$\,. %
Using this fact (together with trivial symmetry of %
the interaction, \,$\Phi(-\rho)= \Phi(\rho)$\,), %
it is easy to proof operator equality
\begin{equation}
\begin{array}{l}
\breve\Gamma(V,\nabla_P)\,G_M(P)\,=\, %
G_M(P)\, \breve{\Gamma}^\ddag(V,\nabla_P)\,=\, %
[\,\breve\Gamma(V,\nabla_P)\,G_M(P)\,]^\ddag\, \,, \,\label{conj}
\end{array}
\end{equation}
where symbol \,$\ddag$\, denotes conjugation %
of operator \,$\breve\Gamma(V,\nabla_P)$\, (or any other %
operator composed of \,$V$\, and \,$\nabla_P=M^{-1}\nabla_V$\,) %
in the Sturm-Liouville sense. %

Taking this into account, one can search for %
eigenfunctions of the RR operator in the form
\begin{equation}
\begin{array}{l}
\breve\Gamma(V,\nabla_P)\,G_M(P)\, %
\Psi_s(V)\,=\, \gamma_s\, G_M(P)\, %
\Psi_s(V)\,\,, \,\,\,\,\,\,  %
\breve\Gamma^\ddag(V,\nabla_P)\,\Psi_s(V)\,=\, %
\gamma_s\, \Psi_s(V)\,\,, \,\label{ef}
\end{array}
\end{equation}
where index \,$s$\, enumerates the eigenfunctions  %
and eigenvalues.
In theur terms, Eq.\ref{dl} turns to %
\begin{equation}
\begin{array}{l}
V_0(t,\Delta R;\nu,f)\,\approx\,
\int_0^\infty %
\left(\frac {\gamma}{4\pi u_0^2 t}\right)^{3/2}\, %
\exp{\left[\,- \frac {\gamma}{4\, u_0^2 t}\, %
\left(\Delta R - \frac {u_0^2}\gamma\, \frac fT\, %
\right)^2\,\right]}\,\, %
W(\gamma)\,d\gamma \,\,,  \, \label{edl}
\end{array}
\end{equation}
with \,$W(\gamma)$\, representing effective RR's probability  %
distribution,
\begin{equation}
\begin{array}{l}
W(\gamma)\,\equiv\, %
\sum_s\, \left|\int \Psi_s(V)\, G_M(P)\, %
dP\,\right|^2\, %
\delta(\gamma -\gamma_s)\,\, \, \label{w}
\end{array}
\end{equation}
Here \,$\sum_s$\, means generally ``continuous summation'', %
and eigenfunctions \,$\,\Psi_s(V)$\, are %
thought properly normalized: %
\begin{equation}
\begin{array}{l}
\int  \Psi^\ddag_{s^\prime}(V)\,  \Psi_s(V)\, %
G_M(P)\, dP\,=\, %
\delta_{s^\prime s}\,\,, \, \label{norm}
\,\,\,\,\,\,
\sum_s\, \Psi_{s}(V^\prime)\,  \Psi^\ddag_s(V)\, =\, %
\delta(P^\prime -P)\, G_M^{-1}(P)\,\,
\end{array}
\end{equation}
Then \,$W(\gamma)$\, also is normalized, %
\,$\int_0^\infty W(\gamma)\,d\gamma =1$\,.

Our presumption that \,$\breve\Gamma$\,'s spectrum %
is real positive is dictated by physical meaning of this %
operator and, of course, by its formal definition %
in Eqs.\ref{rep}-\ref{rep1} and \ref{bgo}  %
prompting that it is positively defined operator:
\begin{equation}
\begin{array}{l}
\int dP\,\, \Psi^\ddag(V)\, \breve\Gamma\,%
\Psi(V)\,G_M(P)\,>\,0\,\, \label{ineq} %
\end{array}
\end{equation}
at nonzero \,$\Psi(V)$\,.

\,\,\,

{\bf 2}.\,\,
Further, notice, or recall, that the (full matrix) RR operator %
is closely connected to the Boltzmann-Lorentz (BL) kinetic operator %
\,$\widehat{\mathcal{B}}= \widehat{\mathcal{B}}(V,\nabla_P)$\,.
According to formulas (57)-(59) from \cite{p0806},
\begin{equation}
\begin{array}{l}
\widehat{\mathcal{B}}\,=\, %
\nabla_P\cdot \breve\Gamma\cdot %
(P+TM \nabla_P)\,\,, \,\, %
\,\,\,\,\,\,
\widehat{\mathcal{B}}G_M\,=\, %
MT\nabla_P\cdot \breve\Gamma G_M\cdot %
\nabla_P\,\,  \, \label{gb}
\end{array}
\end{equation}
This connection, together with fact that spectrum %
of BL operator is non-positive, implies very %
important information about spectrum of %
the RR operator.

Indeed, let us consider expression
\begin{equation}
\begin{array}{l}
C(a,b)\,=\,\int dP\,\, e^{-ia\cdot V}\, %
\widehat{\mathcal{B}}\,e^{\,ib\cdot V}\,G_M(P)\,\, %
\,\,  \label{ca} %
\end{array}
\end{equation}
with real-valued vectors \,$a$\, and \,$b$\,. %
From Eq.\ref{gb} it follows that
\begin{equation}
\begin{array}{l}
C(a,b)\,=\,- \,\frac TM\, a\cdot\int dP\,\, e^{-ia\cdot V}\, %
\breve\Gamma\,e^{\,ib\cdot V}\,G_M(P)\cdot b\, %
\,\,   \label{gba} %
\end{array}
\end{equation}
Next, firstly, take into account that the BL operator %
is non-positively defined, therefore  %
\,$C(a,b=a)$\, is negative quantity at \,$a\neq 0$\,,\, %
\,$C(a,a)<0$\,. %
Secondly, notice that for any short-range repulsive %
interaction this quantity is bounded below,
\begin{equation}
\begin{array}{l}
C(a,a)\,>\,-\gamma^\prime\,\,, \,   \label{bel} %
\end{array}
\end{equation}
 with some finite \,$\gamma^\prime$\,, %
since total collision cross-section is finite. %
As the consequence, Eq.\ref{gba} yields
\begin{equation}
\begin{array}{l}
0\,< \int dP\,\, e^{-ia\cdot V}\, %
\breve\Gamma\,e^{\,ia\cdot V}\,G_M(P)\, %
<\, \frac {\gamma^\prime}{u_0^2\,a^2}\,
\,\,   \label{ineq_} %
\end{array}
\end{equation}

For instance, in case of hard-sphere interaction, %
with radius \,$r_0$\,, one can represent the %
BL operator (in essence integral) %
in differential form
\begin{equation}
\begin{array}{l}
\widehat{\mathcal{B}}G_M\,=\, 2 r_0^2\nu \int \! d\Omega\,  %
\sqrt{\frac {m}{2\pi T}} \int_0^\infty \! du\,\,u\,  %
\exp{\left(-\frac {m^\prime}{2T}\,u^2\right)}\, %
\,\times\\ \,\,\times\,\,
\sinh{(\,qu\,\Omega\cdot\nabla_V)}\, %
G_M(P)\, %
\exp{\left[\,-\frac {m}{2T}\,(\Omega\cdot V)^2\,\right]}\, %
\sinh{(\,qu\,\Omega\cdot\nabla_V)}\,\,, \,\label{bhs} %
\end{array}
\end{equation}
with\, \,$\Omega$\, being unit vector (\,$|\Omega|=1$\,) %
running all over unit sphere, %
and\,  \,$q\equiv m/(m+M)=m^\prime/M$\,\, %
(\,$m^\prime =mM/(m+M)$\,). %
Hence, obviously,  %
\begin{equation}
\begin{array}{l}
C(a,b)\,=\, -\,2 r_0^2\nu \int \! d\Omega\,  %
\sqrt{\frac {m^\prime}{2\pi T}} \int_0^\infty \! du\,\,u\,  %
\exp{\left(-\frac {m^\prime}{2T}\,u^2\right)}\, %
\,\times\\ \,\,\times\,\,
\sin{(\,qu\,\Omega\cdot a)}\, %
\sin{(\,qu\,\Omega\cdot b)}\, %
\exp{\left\{\,-\frac {T}{2M}\, %
\left[\,|a-b|^2\,-\,q\,(\Omega\cdot(a-b))^2\, \right]\right\}}\, %
\,, \,  \label{chs} %
\end{array}
\end{equation}
and\,  \,$\gamma^\prime =4\nu\pi r_0^2\sqrt{T/2\pi m^\prime}$\,. %

The right-hand inequality in Eq.\ref{ineq_} clearly shows %
that lower bound of the RR spectrum, i.e. %
\,$\breve\Gamma\,$\,'s spectrum, is zero:\, %
\begin{equation}
\begin{array}{l}
\inf\, \breve\Gamma\,=\, \inf_s\, %
\gamma_s\,=\,0\, \,\,   \label{inf} %
\end{array}
\end{equation}
(otherwise, on the left here we would have %
some nonzero number in place of zero). %

Simultaneously, \,$\breve\Gamma$\, is an unbounded %
non-compact operator, if being considered  %
in a more wide class of functions than %
in Eqs.\ref{ca}, \ref{gba} and \ref{chs}, namely, a class  %
allowing  \,$|\Psi(V)|\rightarrow\infty$\, %
(polynomially or even exponentially) at \,$|V|\rightarrow\infty$\,.
Therefore \cite{rs} we surely can expect that \,$\breve\Gamma$\,'s %
spectrum is  continuous on \,$(0,\infty)$\,.


{\bf 3}\,\,
The said means, in turn, that BP's diffusion (and drift) law, %
expressed by Eq.\ref{edl}, possesses essentially %
non-Gaussian, - moreover, non-exponential, - long-range %
asymptotic. The latter then is determined by asymptotic of %
the RR distribution \,$W(\gamma)$\, (\ref{w}) at small %
\,$\gamma/\overline\gamma\rightarrow 0$\,. %

Some important statements about this distribution %
and thus diffusion law can be made even %
without detail investigation of the RR operator. %
Anyway, for short-range repulsive interaction,  %
the diffusion law determined by exact basic Eq.{sv} %
must be really ``diffusive'', that is %
undergoe characteristic time scale dependence %
\,$\Delta R^2(t)\propto t$\,, %
at least on average, so that mean value of the diffusivity %
operator in Eq.\ref{do} is finite,
\begin{equation}
\begin{array}{l}
\overline{D}\,\equiv\, u_0^2 \int\! dP\,\, %
\breve\Gamma^{-1} \,G_M(P)\,=\, %
u_0^2\int_0^\infty \frac 1\gamma\, %
W(\gamma)\,d\gamma\,\,<\,\infty\,\,, \,\label{md}
\end{array}
\end{equation}
and thus mean value of linear (low-field) BP's mobility, %
\,$\overline{D}/T$\,, is finite too. %
Using relation (\ref{gb}), we can reformulate this statement %
via the BL operator (BLO), as
\begin{equation}
\begin{array}{l}
\overline{D}\,=\, u_0^2 \int G_M\, %
(\breve\Gamma G_M)^{-1} \,G_M\, dP\,=\,
u_0^2\,TM \int G_M\,\nabla_P\, %
(\widehat{\mathcal{B}} G_M)^{-1}\, %
 \nabla_P\,G_M\, dP\,=\, \,\\ \,\,=\,
-\int G_M\,V\, (\widehat{\mathcal{B}} G_M)^{-1}\, %
 V\,G_M \, dP\, =\, \int V\, %
(-\widehat{\mathcal{B}})^{-1}\, V\,G_M \, dP\, %
<\,\infty \,\, \,\label{md_}
\end{array}
\end{equation}
(under proper treatment of the latter tensor-like expressions). %
Clearly, the ``ground-state'' zero BLO's eigenvalue %
(corresponding to \,$\widehat{\mathcal{B}}G_M =0$\,) %
does not contribute to this formula.
At the same time, any physically meaningful BLO %
has discrete spectrum whose %
non-zero part is separated from zero \cite{rl,rs}. %
Then the inequality in (\ref{md_}) and in (\ref{md}) %
certainly is satisfied, that is the mean diffusivity %
\,$\overline{D}$\, is finite in spite of %
the zero RR's lower bound (\ref{inf}).

Consequently, we can state that RR %
probability distribution vanishes at zero,\, %
\,$W(\gamma\rightarrow 0) \rightarrow 0$\,,\, and %
then guess that it makes this nearly by power law,\,
\begin{equation}
\begin{array}{l}
W(\gamma)\,\propto\, \gamma^\eta\,\,\,\, %
(\,\gamma\,\ll\, \overline\gamma\,)\,\,, \,\,\,\,\,\,
\int_0^\infty e^{ -\gamma t}\, W(\gamma)\, d\gamma\, =\, %
\int\! dP\,\, e^{-t\breve\Gamma} \,G_M(P)\,\propto\, %
\frac 1{t^{\,\eta +1}}\,\,\,\, %
(\,t\,\gg\, 1/\overline\gamma\,)\,\,,  \,\label{eta}
\end{array}
\end{equation}
with\, \,$\eta >0$\,.

Such the asymptotic of RR spectrum, being applied in %
Eq.\ref{dl} or Eq.\ref{edl}, results in strongly %
non-exponential diffusion law at large values of BP's %
displacement (for ``large deviations''). %
In particular, in absence of external force %
(\,$f=0$\,), i.e. in case of %
equilibrium Brownian motion, we have
\begin{equation}
\begin{array}{l}
V(t,\Delta {\bf R}; \nu, f=0)\,\propto\, %
\frac {(\overline{D} t)^{\eta +1}} %
{(\Delta {\bf R}^2)^{\eta +5/2}}\,\,\,\,\,\,\,\,
(\,\Delta {\bf R}^2\,\gg\, \overline{D} t\,)\,\,, \,\label{ld}
\end{array}
\end{equation}
that is power-law long tail for probabilities of %
of large deviations from typical behavior %
(mean-square displacement).

\,\,\, 

{\bf 4}.\,\,
Further, let us consider the RR %
operator (RRO) more carefully, by concrete example %
of the hard-sphere interaction.
In agreement with relation %
(\ref{gb}), RRO what corresponds to the %
hard-sphere BLO (\ref{bhs}) naturally can be %
suggested in the form
\begin{equation}
\begin{array}{l}
(\breve{\Gamma}G_M)_{\alpha\beta}\,=\, %
2 r_0^2\nu \int \! d\Omega\,  %
\sqrt{\frac {m}{2\pi T}} \int_0^\infty \! du\,\,u\,  %
\exp{\left(-\frac {m^\prime}{2T}\,u^2\right)}\, %
\,\times\\ \,\,\times\,\,
\Omega_{\alpha}\, \frac {\sinh{(\,qu\,\Omega\cdot\nabla_V)}}  %
{u_0\,(\Omega\cdot\nabla_V)}\, %
G_M(P)\, %
\exp{\left[\,-\frac {m}{2T}\,(\Omega\cdot V)^2\,\right]}\, %
\frac {\sinh{(\,qu\,\Omega\cdot\nabla_V)}}  %
{u_0\,(\Omega\cdot\nabla_V)}\, %
\Omega_{\beta} \,\,\, \,\label{ghs} %
\end{array}
\end{equation}
It should be noticed that in general so simple formal  %
connection between BLO and RRO has no place, %
moreover, RRO can not be unambiguously extracted from BLO  %
and Eq.\ref{gb}. However, in special hard-sphere case the   %
expression (\ref{ghs}) seems well consistent %
with our basic RRO \,$\breve\Gamma(V,\nabla_P)\,$'s %
definition in (\ref{bgo}).

For more visuality, following the %
scalar RRO's approximation, let us concentrate on %
scalar (mean diagonal) \,$\breve{\Gamma}$\,'s component %
and represent it as evidently integral operator, %
in such way that
\begin{equation}
\begin{array}{l}
\frac 13 \sum_\alpha\, %
G_M^{-1/2}({\bf P})\, %
\breve{\Gamma}_{\alpha\alpha}\, G_M^{1/2}({\bf P})\, %
\Xi({\bf V}) \,=\,
\int \Gamma({\bf V},{\bf V}^\prime)\, %
\Xi({\bf V}^\prime) \, %
d{\bf V}^\prime\,\,, \, \,\label{ie} %
\end{array}
\end{equation}
with definitely positive and symmetric integral kernel %
\,$\Gamma({\bf V},{\bf V}^\prime)=\Gamma({\bf V}^\prime,{\bf V})\,$.
Making in Eq.\ref{ghs} transition to dimensionless velocities, %
\,$V/u_0\Rightarrow V$\, and  %
\,$u/u_0\Rightarrow u$\,, then introducing %
vectorial integration variable \,$qu\Omega $\,, %
and using equality %
\,$\sinh{(x)}/x =\int^1_{-1} \cosh{(cx)}\, %
dc/2$\,, one can transform Eq.\ref{ghs} to
\begin{equation}
\begin{array}{l}
\Gamma({\bf V}_1,{\bf V}_2)\,=\,
\frac {2 r_0^2\nu u_0}{3 q^2}\, %
\sqrt{\frac {r }{2\pi}}\, %
\,\frac 14 \int_{-1}^1 \! \int_{-1}^1 \! %
\frac {dc_1\,dc_2}{(c_1 +c_2)^4}\,\, %
|{\bf U}|\, %
\,\times\\ \,\,\times\,\,
\exp{\left\{\,
-\,\frac {{\bf U}^2}{8q}\cdot %
\frac {4-(c_1-c_2)^2}{(c_1+c_2)^2}\,+\, %
\frac {{\bf U}^2}8\,
%
%
-\, \frac {r }2 \left[\, %
\frac  {({\bf U}\cdot {\bf V})} %
{|{\bf U}|}\,-\, \frac {|{\bf U}|}{2q}\cdot %
\frac {c_1-c_2}{c_1+c_2}\,
\right]^{\,2}\, %
\right\}}\, \,, \,  \label{gk}
\end{array}
\end{equation}
where\, \,\,${\bf U}\equiv {\bf V}_1-{\bf V}_2$\,,\,\,  %
\,${\bf V}\equiv ({\bf V}_1 +{\bf V}_2)/2$\,, %
and \,$r \equiv m/M =q/(1-q)$\,.
Changing of the integration variables,
\[
x\,=\,\frac {4-(c_1-c_2)^2}{(c_1+c_2)^2}\,-\,1\,\,, \,\,\,\,\, %
y\,=\,\frac {c_1-c_2}{c_1+c_2}\,\,, \, 
\]
and then \,$y\Rightarrow 2q y/|{\bf U}|$\,, simplifies this integral to
\begin{equation}
\begin{array}{l}
\Gamma({\bf V}_1,{\bf V}_2)\,=\, %
\frac {\gamma^\prime}{6\pi q\sqrt{1-q}}\cdot %
\frac {|{\bf U}|}{32}\, %
\exp{\left( - \frac {{\bf U}^2}{8r }\right)} %
\, \,\times\\  \,\,\times\, \,\,\,\,\,
\int_{-\infty}^\infty  dy  %
\int_{2|y|}^\infty dx\,\,  %
\exp{\left\{\,-\,x\,\frac {{\bf U}^2}{8q}\, %
-\, \frac {r }2 \left[\, %
\frac  {({\bf U}\cdot {\bf V})} %
{|{\bf U}|}\,-\,y\, \frac {|{\bf U}|}{2q}\,  %
\right]^{\,2}\, \right\}}\,\,  %
= \, \\ \,\,=\,
\frac {\gamma^\prime q}{12\pi \sqrt{1-q}}\cdot %
\frac 1{{\bf U}^2}\, %
\exp{\left( - \frac {{\bf U}^2}{8r }\right)}\, %
\,I({\bf U},\Delta E)\,\,, \,  \,  \label{gk_}
\end{array}
\end{equation}
with\, \,$\Delta E \equiv ({\bf V}\cdot {\bf U}) =  %
{\bf V}_1^2/2 - {\bf V}_2^2/2\,$\, and  function
\begin{equation}
\begin{array}{l}
I({\bf U},\Delta E)\,\equiv\, \int^\infty_{-\infty}
\exp{\left\{\,-\, %
|y|\,\frac {|{\bf U}|}{2}\, -\, \frac {r }2 \left[\, %
\frac  {\Delta E} %
{|{\bf U}|}\,-\,y\,\right]^{\,2}\, \right\}}\, %
dy\, =\, \, \\ \,=\,  %
\sqrt{\frac 2{\pi r }} %
\int_{-\infty}^\infty %
\exp{\left( \,\frac {i\zeta}2\,\Delta E \,-\, %
\frac {\zeta^2{\bf U}^2}{8r }\,\right)}\, %
\frac {d\zeta}{1+\zeta^2}\,\,   \label{iuv}
\end{array}
\end{equation}
It can be exactly expressed via the error function %
(recall that  \,${\bf U}= {\bf V}_1-{\bf V}_2$\,).

These formulas show that at\, %
\,$({\bf V}_1-{\bf V}_2)^2\,\ll\,1$\,\,, i.e. at %
small velocity changes,
\begin{equation}
\begin{array}{l}
\Gamma({\bf V}_1,{\bf V}_2)\,\approx\,  \label{ga}
\frac {\gamma^\prime}3 \sqrt{ \frac q{2\pi}}\cdot  %
\frac 1{({\bf V}_1-{\bf V}_2)^2}\, %
\exp{\left( - \frac {\left|\, \Delta E \,\right|}2\, \right)}\,\,,
\end{array}
\end{equation}
while in the opposite case %
\, \,$\,({\bf V}_1-{\bf V}_2)^2\,\gg\,1\,$\, one has %
\begin{equation}
\begin{array}{l}
\Gamma({\bf V}_1,{\bf V}_2)\,\approx\,  \label{ga_}
\frac {2\gamma^\prime q}{3\pi \sqrt{1-q}}\cdot %
\frac 1{|{\bf V}_1-{\bf V}_2|^3}\,
\exp{\left[\, -\, \frac {|{\bf V}_1-{\bf V}_2|^2}{8r }\,-\,
\frac {r }2\,\left(\, %
\frac {\Delta E} %
{|{\bf V}_1-{\bf V}_2|}\,\right)^2\,\right]}\,\, %
\end{array}
\end{equation}
(recall that\, \,$\Delta E =  %
({\bf V}_1^2 - {\bf V}_2^2)/2\,$\,).

Thus, of course, relaxation rates (RR) for BP's transitions %
between states with nearly equal energies %
(when \,${\bf V}_1^2 \approx {\bf V}_2^2$\, and approximately %
\,${\bf U} \perp {\bf V}$\,) always are greater or much greater %
than RR for transitions between significantly different energies
(when approximately \,${\bf U} \parallel {\bf V}$\,).

\,\,\,

{\bf 5}.\,\, 
The latter statement means that lower part of the %
RR spectrum (\ref{w}) (small \,$\breve\Gamma$\,'s eigenvalues), - %
responsible for long-range asymptotics of diffusion law, - %
is essentially determined by %
\,$\Gamma({\bf V}_1,{\bf V}_2)$\,'s and hence %
\,$I({\bf U},\Delta E)\,$\,'s dependence on %
BP's energy changes\,
\,$\Delta E =({\bf U}\cdot{\bf V}) =  %
({\bf V}_1^2-{\bf V}_2^2)/2\,$.
Especially as, due to obvious  %
spherical symmetry of the operator \,$\breve\Gamma$\,, %
the sum (integral) in Eq.\ref{w} in fact consists of %
its spherically symmetric eigen-states only. %

Therefore, the index \,$s$\, in Eq.\ref{w}  can be treated  %
as one-dimensional continuous parameter, - let \,$\gamma$\,, -    %
varying in \,$0<\gamma <\infty$\, and uniquely %
enumerating non-degenerated spherically symmetric solutions %
of eigenvalue problem
\begin{equation}
\begin{array}{l}
\int \Gamma({\bf V}_1,{\bf V}_2)\, %
\Xi_\gamma({\bf V}^2_2) \, d{\bf V}_2\, =\, %
\gamma\, \Xi_\gamma({\bf V}_1^2)\,\,, \, \label{evp} %
\end{array}
\end{equation}

At that, firstly, if the eigenfunctions %
are normalized to delta-function,\, %
\begin{equation}\begin{array}{l}
\int  \Xi^*_{\gamma^\prime}({\bf V}^2)\,  \Xi_\gamma({\bf V}^2)\, %
d{\bf V}\,=\, 4\pi\int_0^\infty  %
\Xi^*_{\gamma^\prime}(v^2)\,\Xi_\gamma(v^2)\,v^2\,dv\,=\,  %
\delta(\gamma^\prime -\gamma) \,\,, \, \label{nn}
\end{array} \end{equation}
then, - according to Eq.\ref{norm} and obvious   %
relation\, \,$\Psi_\gamma({\bf V})=\Xi_\gamma({\bf V}^2)\, %
G_M^{-1/2}({\bf P})$\,, -
instead of Eq.\ref{w} we have to write
\begin{equation}
\begin{array}{l}
W(\gamma)\,=\, %
\left|\int \Xi_\gamma({\bf V}^2)\, G^{1/2}_M({\bf P})\, %
d{\bf P}\,\right|^2\,=\,
4\sqrt{2\pi} \left|\int_0^\infty v^2\, %
\Xi_\gamma(v^2)\,e^{-v^2/4}\,dv\,\right|^2\, \, \label{ww}
\end{array}
\end{equation}

Secondly, from all the aforesaid it follows that %
lower eigenvalues in Eq.\ref{evp} are contributed %
by more frequently oscillating eigenfunctions.

Thirdly, we beforehand can perform averaging of the %
kernel \,$\Gamma({\bf V}_1,{\bf V}_2)\,$ over %
directions of the velocities. This manipulation, %
with the help of Eq.\ref{iuv}, reduces the problem %
(\ref{evp}) to
\begin{equation}
\begin{array}{l}
\int_0^\infty \overline\Gamma(v_1,v_2) \, %
v_2\, \Xi_\gamma(v^2_2) \, dv_2\, =\,
\int_{-\infty}^\infty
\widetilde{\Gamma}(v_1,v_2) \, %
v_2\, \Xi_\gamma(v^2_2) \, dv_2\, =\, %
\gamma\, v_1\,\Xi_\gamma(v_1^2)\,\,, \, \label{evp_} %
\end{array}
\end{equation}
with\, \,$v_1 \equiv |{\bf V}_1|$\,,\,  %
\,$v_2 \equiv |{\bf V}_2|$\,\,, and functions
\begin{equation}
\begin{array}{l}
\overline\Gamma(v_1,v_2) \,=\, \gamma_0
\int_{(v_1-v_2)^2/r }^{(v_1+v_2)^2/r } %
\frac {dx}x \int_x^\infty \frac {dy}{\sqrt{y}}\,
\exp{\left[\,-\,\frac y8\,-\,%
\frac {(v_1^2-v_2^2)^2}{8y}\,\right]}\, \,=\,\\ \,=\,  \label{gks}
\widetilde{\Gamma}(v_1,v_2)\,-\, %
\widetilde{\Gamma}(v_1,-v_2) \,\,, \, \\ \, %
\widetilde{\Gamma}(v_1,v_2)\,=\, %
\gamma_0\int_{(v_1-v_2)^2/r }^{\infty} %
\ln{\frac {xr }{(v_1-v_2)^2}}\,\,
\exp{\left[-\frac x8- %
\frac {(v_1^2- v_2^2)^2}{8x}\right]}\, %
\frac {dx}{\sqrt{x}}\,
%
\,,
\end{array}
\end{equation}
where\, \,$\gamma_0 \equiv \nu r_0^2 u_0\sqrt{\pi/2}/3$\,.

Notice that the kernel \,$\widetilde{\Gamma}(v_1,v_2)$\, %
has characteristic form
\begin{equation}
\begin{array}{l}
\widetilde{\Gamma}(v_1,v_2)\,=\, %
\gamma_0\,F\left(\frac {v_1-v_2}{\sqrt{r }}\,,\, %
\sqrt{r }\,\,\frac {v_1+v_2}2 \right)\, =\,
\gamma_0\,F^\prime\left(\frac {v_1-v_2}{\sqrt{r }}\,,\, %
\frac {v_1^2-v_2^2}2 \right)\,
\,,   \label{fk}
\end{array}
\end{equation}
with 
functions \,$F(a,b)$\, and \,$F^\prime(a,ab)$\, independent on the %
only RRO's dimensionless free parameter,  %
i.e. mass ratio \,$r =m/M$\,. %
Namely, according to Eq.\ref{gks},
\begin{equation}
\begin{array}{l}
F^\prime(a,ab)\,\equiv\, \int_{a^2}^{\infty} \ln\, \frac {x}{a^2}\,\,
\exp{\left[-\frac x8- \frac {a^2 b^2}{2x}\right]}\, %
\frac {dx}{\sqrt{x}}\, =\,
|a| \int_1^{\infty} \ln{x}\,\, %
\exp{\left[-\frac {x\,a^2}8- \frac {b^2}{2x}\right]}\, %
\frac {dx}{\sqrt{x}}\,\equiv \,F(a,b)\, \,  \label{fk_}
\end{array}
\end{equation}
In other terms, property (\ref{fk}) means that
\begin{equation}
\begin{array}{l}
\widetilde{\Gamma}(v_1,v_2;r )\,=\, %
\widetilde{\Gamma}(v_1\,\cosh{\theta} +  %
v_2\,\sinh{\theta}\,,\, %
v_2\,\cosh{\theta} +  %
v_1\,\sinh{\theta}\,;r _0)\,\,  \label{prop}
\end{array}
\end{equation}
with\, \,$\exp{\theta} \equiv \sqrt{r/r_0}$\,, %
where the kernel's parameter is %
evidently included into list of its arguments.

Approximately, in parallel to (\ref{ga}) and (\ref{ga_}),  %
\begin{equation}
\begin{array}{l}
F(a,b)\,\approx\,\left\{ \begin{array}{c}
\sqrt{2\pi}\,\ln{\frac {4b^2}{a^2}}\, %
\exp{\left(-\frac {|a||b|}2\right)}\, \, %
\,\,\,\,\,\,(\,|a|<2|b|\,)\, \,\,
\\
\ln{\left(1+\frac 8{a^2}\right)}\,\, %
|a| \int_1^\infty \exp{\left(-\frac {a^2x}8\right)}\, %
\frac {dx}{\sqrt{x}}\,\, %
\exp{\left(-\frac {b^2}2\right)}\, \, %
\,\,\,\,\,(\,|a|>2|b|\,)\, %
\end{array} \right\} \,   \label{fa}
\end{array}
\end{equation}
The upper row here comes from the saddle-point %
approximation (applicable at \,$|a|<2|b|$\,).
Its result becomes exact in limit case of infinitely %
massive atoms, \,$r\rightarrow\infty$\,, %
giving
\begin{equation}
\begin{array}{l}
\overline\Gamma(v_1,v_2;\infty)\,=\, %
4\gamma_0 \sqrt{2\pi}\,
\ln{\left|\frac {v_1+v_2}{v_1-v_2}\right|}\,\,
\exp{\left(-\frac  {|v_1^2- v_2^2|}{4}\right)}\, %
\,\,\, \label{ulim}
\end{array}
\end{equation}

\,\,\,

{\bf 6}.\,\,
Unfortunately, these observations not %
 strongly facilitate rigorous analysis  %
of the RR distribution, i.e. calculation of %
quantity (\ref{w}) as determined by equation (\ref{evp_}) %
and condition (\ref{nn}). Therefore   %
it seems reasonable to confine ourselves by    %
semi-heuristic discussion, %
the more so as we are anyway  restricted by our %
initial approximations of the basic path integral.
Then, most important thing is to argue %
that at small RR values, \,$\gamma \ll\overline{\gamma}$\,,  %
the distribution (\ref{ww}), that is (\ref{w}), %
is nearly power function of \,$\gamma$\,,   %
as it was stated in Eq.\ref{eta}.

Indeed, first, according to %
Eqs.\ref{gks},\ref{fk},\ref{fk_},\ref{fa} and \ref{ulim}, the kernel %
\,$\widetilde{\Gamma}(v_1,v_2)=\widetilde{\Gamma}(v_1,v_2;r )$\, %
(even when considered on the whole axis \,$-\infty <v<\infty$\,) %
is not a mere difference kernel  %
but instead a function of two differences  %
\,$v_1-v_2$\,  and \,$v_1^2-v_2^2$\,, both %
playing significant roles.
This means that solutions of the problem (\ref{evp_}), %
\[
v\,\Xi_\gamma(v^2)\,=\,c_\gamma(v)\, %
\sin{\phi_\gamma(v)}\,\,, \,\, 
\]
oscillate in a non-harmonical way, %
that is their phases\, \,$\phi_\gamma(v)$\, %
are essentially non-linear functions of \,$v$\,.
Therefore integral in Eq.\ref{ww} %
is not exponential but (inverse) power functions of %
characteristic frequency of the oscillations %
(for instance,  if \,$\phi_\gamma(v)\propto \zeta v^2$\,, rhen %
this integral \,$\propto 1/\zeta$\, at large enough %
\,$\zeta$\, though it would change like %
\,$\exp{(-\zeta^2)}\,$ if the phase was  %
\,$\phi_\gamma(v)\propto \zeta v$\,).

Second, on the other hand, the characteristic frequency,  %
\,$d\phi_\gamma(v)/dv$\,, is connected with %
the eigenvalues \,$\gamma$\, also by an (inverse) power law, %
merely because of non-analyticity of the kernel   %
\,$\widetilde{\Gamma}(v_1,v_2)$\, at \,$v_1=v_2$\,.
Moreover, since this non-analiticity is dominated by the %
logarithmic factor (obvious in Eqs.\ref{gks},\ref{fk_},\ref{fa} %
and \ref{ulim}), we can state that\, %
\,$d\phi_\gamma(v)/dv \,\propto\, \overline{\gamma}/\gamma$\,.
At that, clearly, since \,$\delta(\gamma^\prime -\gamma) = %
\gamma^{-2} \delta(1/\gamma^\prime -1/\gamma)$\,,  %
 the normalization condition (\ref{nn}) requires %
\,$c_\gamma(v)  \,\propto\, \overline{\gamma}/\gamma$\,,  %
so that \,$v\,\Xi_\gamma(v^2)$\, can be rewritten more visually as
\begin{equation}
\begin{array}{l}
v\,\Xi_\gamma(v^2)\,=\,\frac {c_\gamma(v)}\gamma\, %
\sin{\frac {\phi_\gamma(v)}\gamma}\,\, \,\, \label{sol}
\end{array}
\end{equation}

As the consequence of these two circumstances, (lower part of) RR   %
distribution \,$W(\gamma)$\, (\ref{ww}) %
appears approximately power function of \,$\gamma$\,, %
with exponent \,$\eta =\eta(r)$\, in (\ref{eta}) definitely %
dependent on the mass ratio.

\,\,\,

{\bf 7}.\,\,
In the limit \,$r=\infty$\, %
one may expect this exponent to turn to zero.

To ground this, first, let us represent the eigenfunctions %
by series over normalized Hermite functions associated with %
the equilibrium BP's velocity distribution:
\begin{equation}
\begin{array}{l}
v\,\Xi_\gamma(v^2)\,=\,\sum_{n=0}^\infty\,  %
c_{\gamma\,n}\,\, (-1)^n\, %
\frac {H_{2n+1}(v)}{\sqrt{(2n+1)!}}\, %
\,e^{-v^2/4}\, \,, \,\, \label{solh}
\end{array}
\end{equation}
where \,$H_{k}(x)$\, are standardly defined %
Hermite polynomials. %
Formally such expansion is possible at any \,$r$\,, %
but most adequate it becomes at \,$r=\infty$\, when %
the RRO kernel (\ref{ulim}) directly relates to %
the equilibrium velocity distribution.
Second, compare this series expansion with expression %
(\ref{sol}) and combine both them with equality
\begin{equation}
\begin{array}{l}
8\gamma_0\sqrt{2\pi} \int_0^\infty %
e^{-v^2/4}\, \Xi_\gamma(v^2)\,dv\,=\, %
\gamma\, \Xi_\gamma(0)\,\, \, \label{leq}
\end{array}
\end{equation}
which is easy derivable from Eq.\ref{evp_} in %
the case of kernel (\ref{ulim}).
If we take into account also that
\[
\left(\frac {H_{2n+1}(x)}x \right)_{x=0}\,=\, %
(-1)^n\,(2n+1)!!\,\,, \,\,\,\,\,\,\, %
\int_0^{\infty} \frac {H_{2n+1}(x)}x\, e^{-x^2/2}\, %
dx\,=\, \sqrt{\frac {\pi}2}\, (-1)^n\,2^n\,n!\,\,, \,
\]
and use the Stirling formula, then insertions of  %
(\ref{solh}) and (\ref{sol}) to (\ref{leq}) yield
\begin{equation}
\begin{array}{l}
8\pi\gamma_0 \sum_{n=0}^\infty\,  %
c_{\gamma\,n}\, \left(2\sqrt{\frac %
{n+1}\pi}\right)^{-1/2}\, \approx\,
\gamma \sum_{n=0}^\infty\, c_{\gamma\,n}\,
\left(2\sqrt{\frac {n+1}\pi}\right)^{1/2}\,\, %
\propto \, \frac 1\gamma \,\, \label{ar}
\end{array}
\end{equation}
Here \,$\gamma$\, was presumed so small %
(\,$\gamma\ll 8\pi\gamma_0$\,) that %
$\,\Xi_\gamma(v^2)$\, makes many oscillations per unit.

These relations clearly prompt that at sufficiently small %
\,$\gamma$\, the coefficients \,$c_{\gamma\,n}$\, can %
be approximately expressed through one-variable function, -  %
let \,$c^\prime (\cdot)$\,, - such that
\begin{equation}
\begin{array}{l}
c_{\gamma\,n}\, \propto\, (n+1)^{-1/4}\,  %
c^\prime(\gamma \sqrt{n})\,\,, \,\,\,\,\,\,\,\, %
c^\prime(0)\,=\,1\,\,, \,\, \label{ar_}
\end{array}
\end{equation}
and hence\, \,$c_{\gamma\,n}\,\rightarrow\,$const$\,\neq 0\,$\, %
at \,$\gamma\rightarrow 0$\, for any fixed \,$n$\, %
(numerically, non-zero values of limits %
\,$c_{\gamma\rightarrow 0\,n}\,\,$ are due to unbounded growth of %
the amplitudes in (\ref{sol})).
This behavior of \,$c_{\gamma\,n}$\, %
seems quite natural in view of the fact that %
\,$\sqrt{n}$\, is characteristic frequency of %
\,$H_{2n+1}(x)$\,'s oscillations.

In particular,\, \,$c_{\gamma\rightarrow 0\,0}\,=  %
\int v^2\,\Xi_{\gamma\rightarrow 0}(v^2)\, %
\exp{(-v^2/4)}\,dv\,\neq 0$\,, which means  %
that \,$\eta =0$\,.

\,\,\,

{\bf 8}.\,\,
Of course, more careful consideration of %
the limit case \,$r=\infty$\, could %
reveal some weak, may be logarithmic, %
dependence of \,$\,W(\gamma\rightarrow 0)$\, %
instead of a constant, but this do not cancel equality \,$\eta =0$\, %
in principal sense. Physically, it reflects the fact %
that in gas of infinitely massive %
(hence, immovable) atoms BP's energy does not relax at all.

In the opposite limit of infinitely light atoms, %
\,$r\rightarrow 0$\,, - according to Eqs.\ref{gks}, %
Eq.\ref{fk_} and Eq.\ref{fa}, - the RRO kernel %
effectively reduces to delta-function,
\begin{equation}
\begin{array}{l}
\overline\Gamma(v_1,v_2) \,\rightarrow\, %
32\gamma_0\sqrt{r}\,\delta(v_1-v_2)\, %
\,\,\,\,\,\,\,\, (\,r \rightarrow 0\,)\,\, \, \label{dlim}
\end{array}
\end{equation}
Correspondingly, the RR distribution \,$W(\gamma)$\, %
also shrinks to delta-function, which formally %
means that \,$\eta \rightarrow \infty$\,.
Thus, in this (and only this) special case
one comes to conventional ``Boltzmannian'' kinetics %
with non-random RR.

\,\,\,

{\bf 9}.\,\, 
Notice that, in principle, some properties of the RR operator can %
be investigated even without writing out it evidently, %
instead considering related more usual BL operator.
Indeed, if eigen-states of the full (matrix) RRO, - %
satisfying Eq.\ref{ef}, - can be searched in gradient form %
\,$\Psi_s(V)=\nabla_V \Upsilon_s(V)$\,, then %
application of the divergence operation to %
Eq.\ref{ef} leads vector to scalar problem
\begin{equation}
\begin{array}{l}
\widehat{\mathcal{B}}\,G_M(V) \, %
\Upsilon_s(V)\,=\, \gamma_s\, %
\nabla_V G_M(V) \nabla_V\, %
\Upsilon_s(V)\,\, \,\,\label{ef_}
\end{array}
\end{equation}
(here we applied also the RRO-BLo relation (\ref{gb}), %
but, in contrast to \ref{ef} and (\ref{gb}), take in mind the %
mentioned dimensionless units, in which %
\,$TM=T/M=u_0^2=1$\, and \,${\bf P}={\bf V}$\,).

At that, in order to construct the RR distribution %
(\ref{w}), we can introduce resolvent of the RR operator,
then transforming it as follows:
\begin{equation}
\begin{array}{l}
h(z)\,\equiv\, \int_0^\infty %
\frac {W(\gamma)}{\gamma -z}\, d\gamma\,=\,
\int\! dV\,\, [\,\breve\Gamma -z\,]^{-1} \,G_M(V)\,=\, %
\int G_M\, [\,- zG_M +\breve\Gamma G_M\,]^{-1} \, %
G_M\, dV\,=\, \\ \,=\,
\int G_M\nabla_V\, [\,-z\,\nabla_V G_M\nabla_V + %
\widehat{\mathcal{B}} G_M\,]^{-1}\, \nabla_V G_M\, dV\, %
=\, \\ \,=\,
\int V\, [ \,z\, \nabla_V (V +\nabla_V) - %
\widehat{\mathcal{B}} \,]^{-1}\, V\, %
G_M\, dV\,\,  \,\label{res}
\end{array}
\end{equation}
Here again formal transformations %
like that in (\ref{md_}) are performed, - thus %
visually connecting RRO resolvent to the BLO, - %
and again BLO's zero eigen-states do not contribute  %
to result. Formally, if knowing (\ref{res}), one can %
restore RR distribution with the help of
\[
W(\gamma)\,=\, \frac 1\pi\, \Im\, %
\lim_{\epsilon\rightarrow 0}\, %
h(\gamma +i\epsilon)\,
\]

The last expression in Eq.\ref{res}  %
differs from usual resolvent by operator-valued argument %
\,$z\nabla_V(V + \nabla_V)$\, in place of a %
(complex) {\it c}-number valued argument like \,$z$\,.
Thus, RR statistics can  be determined from spectral properties %
of linear combination of two operators, namely, BLO  %
and simplest Fokker-Planck operator, \,$\nabla_V (V +\nabla_V)$\,.
Each of them has (non-positive) countable  %
spectrum. Nevertheless, as far as they do not commute %
one with another, their linear combination can possess  %
continuous spectrum, as it was shown above.

\section{Conclusion}

\[
\begin{array}{r}
\,\,\,\,\,\,\,\,\,\,\,\, \texttt{"Get at the root!"} \,\,\,\,\,
\textit{(Koz'ma Prutkov)}\,\,\,
\end{array}
\]

\,\,\,

{\bf 1}.\,\,
We considered the original path integral which was    %
derived in \cite{p0806} from complete infinite %
hierarchy of the BBGKY (Bogolyubov-Born-Green-Kirkwood-Yvon) %
equations as exact representation for non-equilibrium %
probability distribution of trajectory (path) of a particle  %
interacting with atoms of equilibrium ideal gas %
\cite{p0806,ig,hs,p1209,p0802,p0803,p0804,tmf}.
Since this is extremely non-Gaussian path integral, %
our main purpose here was, firstly, to recognize a non-trivial  %
but calculable approximation for it %
and, secondly, demonstrate that corresponding results %
are in natural agreement with results of our earlier %
investigated approximate approaches to %
solution of the BBGKY equations  %
\cite{i1,i2,p1,p1007,p1105,p1209}.

Both these objects are achieved by means of %
approximation which formally separates %
two floors of randomness in walk %
of the ``Brownian'' particle under %
consideration (BP), namely, its directly  %
obsevable velocity fluctuations %
and less visible deeper hidden fluctuations %
im their intensity,  %
that is fluctuations in rate of velocity relaxation and agitation %
because of the BP's collisions with gas atoms.

Importantly, such approximation  %
appears next after most primitive one which %
neglects the relaxation rate (RR) fluctuations at all  %
and therefore replaces actual path  %
integral by a Gaussian integral while true BP's %
random walk by Ornstein-Uhlenbeck random process  %
obeying a Fokker-Planck kinetic equation.
What is for the Boltzmann-Lorentz equation, %
there are no evident conditions for its appearance %
in some approximation. This fact once more  %
confirms our statement, - for the first time deduced %
in \cite{i1} and especially grounded in %
\cite{hs1,p1203}, - that the Boltzmann kinetic equation is %
invalid even for arbitrary dilute gases and hence  %
the Boltzmann-Lorentz (BL) equation is invalid too.

\,\,\,

{\bf 2}.\,\,
Dramatic defect of Boltzmannian kinetics, - %
explained in detail already in \cite{i1}, in principle %
even earlier in \cite{pr157,bk1,bk2,ufn,pr195} and %
additionally in other  words in   %
\cite{ig,i2,i3,p1,p0710,p0802,tmf,kmg,e1,e2,p1203,yuk,ufn1}, -
is that it always unconsciously postulates %
constant ``probabilities of collisions %
per unit time'', or ``collision cross-sections'', etc., %
in place of actually random quantities which have no tendency %
to ``time self-averaging''.

For example, relative frequency of BP's collisions %
of a given sort (e.g. with given impact parameter value)  %
has no definite time average, merely because its %
random deviations from an imaginary ``norm''  %
(e.g. theoretical ensemble average) %
do not cause a compensating ``back reaction''  %
of the system, instead being constantly %
forgotten by it.
Therefore, a hystogram of collisions' distribution %
over sorts (impact parameters), - as well as summary   %
rate of various collisions, - stay not %
smoothed out with time, instead undergoing time-scaleless %
(1/f-type) fluctuations irregularly %
distributed over the hystogram \cite{i1,i2}.

In essence, all that was proved already in  %
\cite{kr} (see Introduction) as logically %
inevitable consequence of the %
mixing property (exponential instability %
and chaoticity) of many-particle dynamics.

This dictates quite clear logics:\, %
if a flow of events produced by dynamical %
system is memoryless, - that is its past history does not  %
influence onto its future, - then such flow has no %
certain ({\it \,a priori\,} predictable) ``probabilities of %
events per unit time''  %
\footnote{\,  %
In other words, `probabilities per unit time'' (or other %
characteristics with similar meaning) are different at %
different phase trajectorie of a system in theory and %
in different experiments in practice.}\,. %
Besides, also logically inevitably, in such purely random flow %
all events are mutually statistically correlated, %
since all are commonly and equally responsible for resulting %
({\it \,a posteriori\,}) number of events per unit time.
Thus, as it was underlined in \cite{kr}, %
in statistical mechanics %
physical (cause-and-consequence) independence of events %
does not necessarily mean their statistical independence %
\footnote{\,  %
To avoid confusions with folklore %
``ergodic hypotheses'' and ``ergodic theorems'', %
one should identify the  %
``events'' with transitions between instant system's  %
states rather than states themselves.
On true relations between uncertainty of %
``probabilities'' of events and 1/f-noise, on one hand, %
and the ergodic theory, on the other hand,  %
see e.g. Introduction in \cite{tmf}.}\,. %

In practice, unfortunately, scientists exploit %
the inverse ``medieval'' %
Bernoulli's \cite{jb} logics, thinking that physically %
independent events must be statistically independent  %
and furnished with {\it \,a priori\,} definite %
personal probabilities.
An user of these assumptions and thus of %
the Bernoulli's ``law of large %
numbers'' \cite{jb} implied by them, -  %
e.g. in Boltzmannian kinetics, -  %
usually is not aware that they %
automatically kill many potential possibilities %
for statistics of random events to  %
be described  %
\footnote{\, %
Really, the law of large numbers shows  %
that assigning personal independent probabilities %
to events is nothing but assigning {\it \,a priori\,}  %
certain time limit to ``number of events per unit time''.
Moreover, even  assigning {\it \,a priori\,} certain  %
total number of events (since its relative %
uncertainty fast enough tends to zero).
From physical viewpoint, this looks very strangely, %
as if all events were obeying some  %
teleological external control and/or %
infinitely long-living common memory. %
Anyway, any low-frequency fluctuations {\it \,a priori\,} %
are excluded from such a theory.}\,.

Therefore, it will be not surprising  %
if in this way of thinking the problem of ``ubiquitous %
1/f noise'' never will be resolved  %
\footnote{\, %
On the contrary, in such way the beautiful mathematical %
probability theory may imperceptibly become turned into %
a kind of physical pseudo-science (a good example may be %
the factually axiomatic construction of hard-sphere gas %
kinetics, - see e.g. references and remarks in %
\cite{p0806,ig,hs,tmf,hs1,p1203}). Though, of course, %
{\it \,homo erratum est\,}, and modern scientific ``homo'' not less %
than antique \cite{zin}. And, on the other hand, the firth %
Euclid axiom equally can be accepted or refused.}\,.  %

\,\,\,

{\bf 3}.\,\,
These remarks may help to understand meaning of  %
our path integral and its approximation  %
corresponding to the ``two-storey'' division of %
BP's motion into velocity fluctuations, %
Gaussian in themselves, and non-Gaussian %
fluctuations of velocity relaxation rate (RR).  %
We named this ``free RR approximation'',  %
since in it the lower storey (RR fluctuations) %
keeps independent on the upper one (i.e. on resulting BP's path).  %

Equivalently, one can %
speak about fluctuations of velocity relaxation time, %
or spectral intensity (power density), %
or BP's diffusivity and mobility,  %
or relative frequency and efficiency of BP's collisions with atoms,  %
or their effective cross-section, etc.

In the path integral approach all that  %
is contained in any of several forms of the ``interaction kernel'',  %
particularly, in functional \,$\Gamma\{t,{\bf A}^*,{\bf A}\}$\,  %
and equivalent RR operators %
\,$\widehat{\Gamma}(A^\dagger,A)$\, and  %
\,$\breve{\Gamma}(V,\nabla_P)$\, or, equivalently, %
in their integral-operator kernel \,$\Gamma({\bf V}_1,{\bf V}_2)$\,,  %
all introduced in Sections 7-8 for (most interesting for us here)  %
the case of dilute gas and short-range BP-atom interaction.

Applying quantum terminology, these objects are %
analogues of ``scattering matrix'', or  %
``scattering operator'', of BP in gas. %
However, unlike usual kinetics,  %
in our theory not matrix elements of this operator  %
but its eigen-values and eigen-functions are of most importance. %
The eigen-values represent possible %
values of global, - that is  {\bf time-averaged}, - %
RR of BP's velocity fluctuations.
At that, spectrum of the eigenvalues never reduces to %
a single point (except may be limits of infinitely %
light atoms or infinitely hard BP). %
In opposite, we have shown that it occupies %
the whole real positive semi-axis (hence %
including arbitrary close vicinity of zero). %

This principal mathematical fact means that %
BP's random walk always possesses %
``kinetic non-ergodicity'', i.e. different %
relative frequencies and efficiencies of collisions %
(and, as a result, different diffusivities and mobilities)  %
at different observations (experiments).

Thus, in the present paper for the first time   %
a mathematical object - the RR operator - %
is pointed out what serves, under definite %
approximation, as quantitative measure of the %
kinetic non-ergodicity, being evidently expressed  %
in terms of thr underlying interaction potential and %
collision dynamics.

Inter-relation between eigen-functions   %
of this operator and its above mentioned eigen-values says that %
the more irregular (sharply oscillating) %
is a perturbation of initial BP's velocity %
probability distribution (histogram)  %
the longer is its relaxation, in the sense that it weaker  %
affects diversity of following BP's trajectories %
and diffusivities exposed on them. %
In other words, in any set of BP copies (with  %
randomly chosen initial states) all they  %
almost surely will demonstrate substantially different %
diffusivities, - i.e. spectral powers of velocity %
fluctuations, - regardless of duration of their measurements.
This, of course, is mere manifestation of exponential  %
instability of dynamical BP's path in respect to its perturbations.

\,\,\,

{\bf 4}.\,\,
At the same time, the RR operator is not something  %
unprecedented. In fact, as was shown, %
at least in the case of dilute gas, it is simply   %
connected with well known object, %
namely, the standard Boltzmann-Lorentz (BL) operator %
(on it see e.g. \cite{rl}).
More concretely, the RR operator (RRO) is %
sealed up inside the BL operator (BLO) as its %
constituent part, like ``genie in bottle''.  %

There, inside the BLO, the RRO can play %
significant role by introducing quantitative dependence of %
BLO spectrum on the atom-BP mass ratio \,$r=m/M$\,. %
But qualitatively, - from viewpoint of %
long-range statistics of BP's random walk, - %
its presence there is not important. %
Indeed, the same (standard Gaussian) asymptotic diffusion law %
would take place even if we substituted the RRO inside BLO  %
for trivial {\,c}-number constant, that is turned the BLO %
into Fokker-Planck operator %
\footnote{\, %
In connection with this remark, it is interesting   %
to recollect that the theory based on %
the BBGKY equations  \cite{i1,i2,p1,p1007} %
does not lose the diffusivity 1/f noise (and related  %
non-Gaussianity of diffusion law) if it takes into account  %
conservation of center-of-mass velocity in every %
particular collision (and thus in any many-particle %
cluster of collisions), even though approximating %
momentum and energy exchange between colliding particles %
by artificial model Fokker-Planck collision operator.  %
This fact highlights deep physical meaning of arbitrary  %
small RRO's eigen-values:\,  %
they delegate conservation of total momentum of %
BP plus gas during collisions.  %

In this respect, notice that significant part of BP's path  %
always is collective center-of-mass flight %
whose contribution can be expressed by\, %
\,$t\,[MV+m\sum_{j=1}^{N(t)} v_j]/  %
[M+mN(t)]$\,, where \,$N(t)$\, denotes number of past BP's %
collisions with some of atoms during observation time %
 \,$t$\,, while \,$V$\,.and \,$v_j$\, present %
 (or initial) velocities of BP and that atoms.
}\,. %

But the ``release of the genie'' in exact theory %
(or let approximate but correct enough in %
comparison with naive Boltzmannian kinetics)   %
radically ``changes things around''.
Now, the mass ratio enters not only %
into mean BP's diffusivity (mobility), but  %
also into forms of RR and diffusivity (mobility)  %
probability distributions and, consequently, into  %
very qualitative construction  %
and visual shape of (now non-Gaussian) diffusion law  %
(total BP's path probability distribution).

In particular,\, \,$r=m/M$\, determines %
amplitude and exponent of power-law long tails  %
of long-time asymptotic of diffusion law, %
which in turn are closely connected to magnitude  %
and spectrum exponent of low-frequency diffusivity  %
(mobility) 1/f noise \cite{i1,p1,p0802,tmf,p1007,p1209}.

Unfortunately, calculation of  %
the RR distribution was out of our possibilities here, %
since it requires accurate analysis of  %
structure of the RRO eigen-functions (if not exact %
diagonalization of RRO). %
Nevertheless, for principal comparison %
with earlier results even our %
rough discussion of this distribution is sufficient %
(especially if remembering that the RRO itself %
is product of approximation of exact path integral).
The discussion prompted that the RR distribution  %
exponent \,$\eta$\, is nearly inversely proportional to the  %
mass ratio,\, \,$\eta(r)\propto 1/r$\,. %
This is in agreement with results obtained in %
the frameworks of two quite different approaches, - %
the ``collisional approximation'' to the BBGKY %
hierarchy \cite{i1,i2,p1,p1007}   %
and the rigorous ``dynamical virial relations'' %
\cite{p0710,p0802,p0803,tmf,p1105,p1209}, -  %
both yielding \,$\eta(1)= 1$\, \cite{p1,p0802}  %
\,$\eta(r)= 1/r$\, \cite{p1007,p1209}.

Hence, we can state that now three very different  %
approaches to the problem give essentially similar %
results, thus justifying one another.

\,\,\,

{\bf 5}.\,\,
Our consideration allows to expect that the expounded %
(here and in \cite{p0806}) pseudo-quantum path integral %
representation of classical statistical dynamics  %
can be usefully extended to ``Brownian motion'' in non-ideal gases  %
(fluids) with arbitrary density. Moreover, to %
mutual diffusion and drift and other transport processes %
in mixtures of different fluids, and may be even %
to the hydrodynamics as the whole.

A principal base for such the extension is %
given by the Bogolyubov functional evolution  %
equation \cite{bog} which, however, %
preliminarily should be reformulated  in terms %
of suitably defined irreducible many-particle  %
correlation (cumulant) functions and %
their generating functionals, %
in analogy with what was made in  %
\cite{p0806,p0705,p0710,p0802,p0803,tmf,p1105,p1203,p1209}.
Then, after such preparation, one can introduce %
effective quantum field operators like \,$a^\dagger(x)$\,  %
and \,$a(x)=\delta/\delta a^\dagger(x)$\, above, %
with \,$x$\, enumerating \,$\mu\,$-phase space  %
points (and fluid species).

Of course, on this way generally we have to deal with more %
complicated effective ``Hamiltonians'' %
(in essence, Liouville operators) than previously %
met ones.  At least, with Hamiltonians containing %
four-particle interactions, i.e. terms with  %
\,$a^\dagger(x)a^\dagger(y)\,a(x^\prime)\,a(y^\prime)$\,.
Nevertheless, the corresponding formalism, including  %
path integrals, undoubtedly will be able to create  %
meaningful non-trivial approximations  %
(and, may be, unify the pseudo-quantum and truly quantum %
theories  %
\footnote{\,  %
Notice that in \cite{e1,e2} very close %
formalisms for actually quantum Brownian particles in  %
quantum many-particle thermostats already were suggested and  %
partly developed (up to rather general ``theorem %
on fundamental 1/ noise'' \cite{p1207}).}\,).  %
I wish good luck to followers.



\end{document}